\documentclass[twocolumn]{aastex631}

\usepackage{subfigure}
\usepackage{epsfig}
\usepackage{longtable}
\usepackage{isotope}
\usepackage{threeparttable}
\usepackage{nicematrix,tikz}
\usepackage{array}
\usepackage{booktabs}
\usepackage{threeparttable}
\usepackage{tabularx}
\usepackage{multirow}

\begin{document}

% \title{Arcs from the weak: a search of strongly lensed faint dusty star-forming galaxies in COSMOS-Webb}
\title{Unveiling a Population of Strong Galaxy-Galaxy Lensed Faint Dusty Star-Forming Galaxies}

\correspondingauthor{Ting-Kai Yang, Chian-Chou Chen}
\email{tkyang@asiaa.sinica.edu.tw, ccchen@asiaa.sinica.edu.tw}

\author[0009-0000-8958-3780]{Ting-Kai Yang}
\affiliation{Department of Physics, National Taiwan University, Taipei 106319, Taiwan}
\affiliation{Academia Sinica Institute of Astronomy and Astrophysics (ASIAA), 
11F of Astronomy-Mathematics Building, AS/NTU, No. 1, Section 4, 12 Roosevelt Road, Taipei
106319, Taiwan}

\author[0000-0002-3805-0789]{Chian-Chou Chen}
\affiliation{Academia Sinica Institute of Astronomy and Astrophysics (ASIAA), 
11F of Astronomy-Mathematics Building, AS/NTU, No. 1, Section 4, 12 Roosevelt Road, Taipei
106319, Taiwan}

\author[0000-0003-1262-7719]{Zhen-Kai Gao}
\affiliation{Academia Sinica Institute of Astronomy and Astrophysics (ASIAA), 
11F of Astronomy-Mathematics Building, AS/NTU, No. 1, Section 4, 12 Roosevelt Road, Taipei
106319, Taiwan}

\author[0000-0003-2213-7983]{Bovornpratch Vijarnwannaluk}
\affiliation{Academia Sinica Institute of Astronomy and Astrophysics (ASIAA), 
11F of Astronomy-Mathematics Building, AS/NTU, No. 1, Section 4, 12 Roosevelt Road, Taipei
106319, Taiwan}

\author[0000-0001-9882-1576]{Adarsh Ranjan}
\affiliation{Space Telescope Science Institute, 3700 San Martin Drive, Baltimore, MD 21218, USA}

\author[0000-0003-2588-1265]{Wei-Hao Wang}
\affiliation{Academia Sinica Institute of Astronomy and Astrophysics (ASIAA), 
11F of Astronomy-Mathematics Building, AS/NTU, No. 1, Section 4, 12 Roosevelt Road, Taipei
106319, Taiwan}

\author[0000-0002-0930-6466]{Caitlin M. Casey}
\affiliation{Department of Physics, University of California, Santa Barbara, Santa Barbara, CA 93109, USA}
\affiliation{The University of Texas at Austin, 2515 Speedway Blvd Stop C1400, Austin, TX 78712, USA}
\affiliation{Cosmic Dawn Center (DAWN), Denmark}

\author[0000-0002-6821-8669]{Tomotsugu Goto}
\affiliation{Institute of Astronomy, National Tsing Hua University, 101, Section 2. Kuang-Fu Road, Hsinchu, 30013}

\author[0000-0001-9187-3605]{Jeyhan S. Kartaltepe}
\affiliation{Laboratory for Multiwavelength Astrophysics, School of Physics and Astronomy, Rochester Institute of Technology, 84 Lomb Memorial Drive, Rochester, NY 14623, USA}

\author[0000-0003-4531-0945]{Chayan Mondal}
\affiliation{Academia Sinica Institute of Astronomy and Astrophysics (ASIAA), 
11F of Astronomy-Mathematics Building, AS/NTU, No. 1, Section 4, 12 Roosevelt Road, Taipei
106319, Taiwan}

\author[0000-0001-8555-8561]{James Pearson}
\affiliation{School of Physical Sciences, The Open University, Milton Keynes, MK7 6AA, UK}

\author[0000-0002-4375-5978]{Chris Sedgwick}
\affiliation{School of Physical Sciences, The Open University, Milton Keynes, MK7 6AA, UK}

\author[0000-0002-0517-7943]{Stephen Serjeant}
\affiliation{School of Physical Sciences, The Open University, Milton Keynes, MK7 6AA, UK}

% \collaboration{20}{(AAS Journals Data Editors)}

% \author{F.X Timmes}
% \affiliation{Arizona State University}
% \affiliation{AAS Journals Associate Editor-in-Chief}

% \author{Amy Hendrickson}
% \altaffiliation{AASTeX v6+ programmer}
% \affiliation{TeXnology Inc.}

% \author{Julie Steffen}
% \affiliation{AAS Director of Publishing}
% \affiliation{American Astronomical Society \\
% 1667 K Street NW, Suite 800 \\
% Washington, DC 20006, USA}



\begin{abstract}
The measurements of the number density of galaxy-galaxy strong lenses can be used to put statistical constraints on the foreground mass distributions. Dusty galaxies uncovered in submillimeter surveys are particularly useful in this regard because of the large volume probed by these surveys. Previous discoveries of strong galaxy-galaxy lensed dusty galaxies are predominantly the brightest in the sky discovered by Herschel, SPT, and Planck. However, models have also predicted a non-negligible fraction of strong galaxy-galaxy lensed faint dusty galaxies, which were difficult to confirm due to technical difficulties. Utilizing the deepest SCUBA-2 submillimeter survey, STUDIES, in both the COSMOS and the UDS fields, together with a red JWST color selection method, we discover a population of 13 strong galaxy-galaxy lensed faint dusty galaxies. The rich ancillary data allow us to confirm their strongly lensed nature via estimates of redshifts and lens modeling. Our systematic search has allowed us to construct the 450\,$\micron$ number counts of strongly lensed sources down to the flux levels about an order of magnitude fainter than previous measurements. The measured lensing fractions of $\sim$1\% are consistent with predictions from models that also successfully produce the number density of the strong galaxy-galaxy lensed bright dusty galaxies. Future searches from Euclid and Roman are expected to discover orders of magnitude more strongly lensed faint dusty galaxies.

\end{abstract}



\section{Introduction} \label{sec:intro}

%% A general introduction on how measuring the fraction of sources that are strongly lensed can help put constraints on matter distributions.

%% Why dusty galaxies are particularly useful.

%% A brief history of what has been found for strongly lensed dusty galaxies. Mostly bright sources from the SPT and Herschel surveys. Successful in putting constraints on the brightest end of the strongly lensed number densities. 

%% With the advent of deep surveys and JWST, the number density of strongly lensed faint dusty galaxies can start to be measured. Indeed, recent report of individual lensed sources (Pearson et al. 2024; van Dokkum et al. 2024) have demonstrated the feasibility. 

%% Here we perform a systematic search of the strongly lensed faint dusty galaxies.

Strong galaxy-galaxy lensing is a powerful tool to probe dark matter by analyzing the foreground gravitational potential responsible for the lensing effect. Fluctuations in mass density along the line of sight significantly impact the gravitational lensing effect, and the spatial mass distribution of the lenses related to the content of dark matter and dark energy \citep{Kochanek1992,Mitchell2005,Grillo2008,Eales2015,Shajib2024}. The number density of the strongly lensed sources hence become a crucial measurement in studying mass distribution in the Universe, as the detection of more strongly lensed galaxies implies a greater amount of foreground mass. Apart from studying the mass distribution, galaxy lensing studies also play a crucial role in other areas of research. For example, high-resolution data allows for the detailed examination of the morphology and substructure within the foreground lens galaxies \citep{Mao1998,Dalal2002,Vegetti2010,Hezaveh2016,Powell2023}. %By investigating the mass distribution of these lenses, we can gain deeper insights into the nature of dark matter and test various dark matter scenarios  
%The abundance of lensed galaxies allows constraints to be placed not only on the mass distribution but also on cosmological parameters, such as the mass density parameter. 
The Hubble constant $H_0$ can be independently constrained through time-delay measurements of strongly lensed quasars or supernovae \citep{Refsdal1964,Suyu2010,Wong2020,Treu2022,Kelly2023}. The observations taken on the strongly amplified and stretched background lensed galaxies also allow investigations of galaxy properties to exquisite details that are otherwise difficult \citep{Muzzin2012,ALMAPartnership2015,Johnson2017,Canameras2017,Spilker2022}.

Dusty star-forming galaxies (DSFGs) are characterized by their high star formation rate (SFR) and high luminosity in the far-infrared and submillimeter wavelengths \citep{Blain2002,2014PhR...541...45C}. These galaxies represent an essential population in the galaxy evolution with extreme star-forming activity. Since most of the DSFGs are located at cosmic noon ($z\sim2$) and beyond \citep{Chapman2005,Dudzeviciute2020,Lim2020,Chen2022}, together with the large volume probed by submillimeter surveys thanks to the negative $k$-correction, DSFGs are particularly useful for constraining the foreground mass distribution with a higher probability aligning with other foreground galaxies along the line of sight. 
% In addition, detailed studies of the DSFGs are conducted through gravitational lensing, which magnifies both their angular size and luminosity. Identifying these strongly lensed DSFGs is therefore essential for expanding the sample sizes needed for statistical analyses, helping to unravel the physical properties and evolutionary role of these galaxies. 

In previous searches of the strongly lensed DSFGs, these sources have predominantly been found in wide and shallow submillimeter or millimeter surveys. In the Herschel Astrophysical Terahertz Large Area Survey (H-ATLAS), the largest submillimeter survey conducted by Herschel, a population of strongly lensed DSFG was identified with flux-limited selection (e.g., $S_{500\,\micron}\gtrsim80-100$\,mJy; \citealt{2010Sci...330..800N, 2017MNRAS.465.3558N, 2018MNRAS.473.1751B, 2022MNRAS.511.3017U}). 
% With the model prediction of nearly no un-lensed DSFG could reach the submillimeter brightness, the strongly lensed DSFGs found in H-ATLAS were identified by applying flux cut in submillimeter wavelength. 
The same method was used to search for candidates in other Herschel surveys \citep{Wardlow2013,2016ApJ...823...17N}, as well as surveys from other facilities, such as the South Pole Telescope (SPT; \citealt{2013Natur.495..344V,2020ApJ...902...78R}) and Planck \citep{2016MNRAS.458.4383H, 2021A&A...653A.151T}.
% South Pole Telescope (SPT) also observed a population of DSFGs with excessed submillimeter or millimeter brightness, which were later confirmed to be caused by strong lensing \citep{2013Natur.495..344V,2016ApJ...826..112S,2020ApJ...902...78R} and widely used for investigating the physical properties of DSFGs \citep{2023ApJ...948...44R}. The strongly lensed DSFGs were also detected in Planck submillimeter survey \citep{2016MNRAS.458.4383H, 2021A&A...653A.151T}, which is shallower than H-ATLAS but covers all-sky range, giving a larger sample of strongly lensed DSFGs selected with submillimeter flux criteria for statistical studies. 
% With the submillimeter- or millimeter-bright strongly lensed sources found, 
These surveys successfully put measurement constraints on the bright end of the strongly lensed DSFG number density. However, the constraints at the faint end are lacking, where the models have predicted that the fractions of strongly lensed sources to be at the level of $\sim1$\% \citep{Cai2013,2017MNRAS.465.3558N}.

Recently, a few individual cases of strongly lensed faint DSFGs have started to be discovered thanks to the advent of deep surveys conducted by the James Webb Space Telescope (JWST; \citealt{2024NatAs...8..119V, 2024MNRAS.52712044P, 2024Mercier}). 
% With the high-resolution data in near infrared, the strongly lensed but submillimeter-faint dusty galaxies can be identified from the distorted morphology.
% The recent report of individual lensed sources \citep{2024NatAs...8..119V, 2024MNRAS.52712044P, 2024A&A...687A..61M} have demonstrated the feasibility. 
To put constraints on the strongly lensed DSFG number density at the faint end, here we perform a systematic search based on the ultra-deep SCUBA-2 survey STUDIES \citep{2017ApJ...850...37W}, as well as recently established color selection methods.

In Section~\ref{sec:obs} we describe the data we used. In Section~\ref{sec:analyses} we present our search and the results. In Section~\ref{sec:sum} we present the summary. We assume flat $\Lambda$CDM cosmology with $H_0=70$ km s$^{-1}$Mpc$^{-1}$, $\Omega_M=0.3$ and $\Omega_\Lambda=0.7$ throughout this paper.

\section{Observations and Data} \label{sec:obs}

\subsection{Infrared Data}\label{sec:infrared}

The JWST imaging data used in this work are based on Near Infrared Camera (NIRCam) observations taken by the PRIMER \citep{2021jwst.prop.1837D} and COSMOS-Web surveys \citep{2023ApJ...954...31C}. All the JWST data used in this paper can be found in MAST: \dataset[10.17909/1nkf-fn20]{http://dx.doi.org/10.17909/1nkf-fn20}. COSMOS and UDS field are targeted by PRIMER, using 8 bands NIRCam filters (F090W, F115W, F150W, F200W, F277W, F356W, F410M, and F444W), as well as two bands (F770W and F1800W) from the Mid-Infrared Instrument (MIRI). The total area PRIMER covered in COSMOS and UDS field is 144.72 square arcminutes and 255.88 square arcminutes, respectively. The NIRCam footprint of COSMOS-Web covers $\sim$0.54 square degrees in four bands (F115W, F150W, F277W, and F444W) with a 5$\sigma$ depth reaching down to around 27 magnitudes.

The JWST data products used in this study were processed using a customized data reduction pipeline, built upon the official pipeline (v1.13.4) within the context of \texttt{jwst\_1237.pmap}. In addition to the official pipeline, our Stage 1 processing includes a snowball masking step, implemented with \texttt{snowblind}\footnote{https://github.com/mpi-astronomy/snowblind/}. In Stage 3, we introduce several additional steps to mitigate 1/f noise, correct for pedestals and artifacts, address guide star failures that lead to inaccurate world coordinate system (WCS) information, and optimize the outlier detection and mosaicking processes using algorithms specifically designed for large mosaics. For further details, please refer to Vijarnwannaluk et al. (in preparation) for NIRCam reduction and Gao et al. (in preparation) for MIRI reduction.

To facilitate the search of the red DSFGs, in this work, two catalogs derived from the NIRCam imaging data were used. The first catalog, referred to as the original catalog, contains sources extracted from the calibrated NIRCam images in the F277W and F444W filters, with flux values corrected from aperture to total. The second catalog, dubbed the threshold catalog, was specifically created to detect only redder sources and was made as the following. The F444W-F150W difference map was first generated by subtracting the F150W flux from the F444W flux. 
% Since the dusty and red background sources may be faint and blended with the bluer foreground in shorter wavelengths, 
% we identified background sources by selecting red objects from a difference map between the long and short wavelengths. 
% the threshold catalog optimizes the identifications of the red lensed sources, as well as the creation of source aperture for photometric measurements. 
% , resulting in the F444W-F150W difference map. 
Sources with a detection above 10$\sigma$ in this difference map were extracted to form the threshold catalog. This method effectively removed blue objects, producing a catalog optimized for searching red and dusty galaxies. In this threshold catalog, isophotal fluxes across multiple bands were measured in the same segmentation made in the detection process, providing more accurate color information for lensed and extended sources.

\begin{deluxetable*}{lccccccccc}[ht]
%% Keep a portrait orientation

%% Over-ride the default font size
%% Use Default (12pt)
%% Use \tablewidth{?pt} to over-ride the default table width.
%% If you are unhappy with the default look at the end of the
%% *.log file to see what the default was set at before adjusting
%% this value.

%% This is the title of the table.
\tablecaption{Strongly lensed DSFG candidates in COSMOS field.}
%% This command over-rides LaTeX's natural table count
%% and replaces it with this number.  LaTeX will increment 
%% all other tables after this table based on this number
\tablenum{1}
%% The \tablehead gives provides the column headers.  It
%% is currently set up so that the column labels are on the
%% top line and the units surrounded by ()s are in the 
%% bottom line.  You may add more header information by writing
%% another line between these lines. For each column that requries
%% extra information be sure to include a \colhead{text} command
%% and remember to end any extra lines with \\ and include the 
%% correct number of &s.
\tablehead{\colhead{ID\tablenotemark{a}} & \colhead{RA} & \colhead{Dec} & \colhead{$z_\text{spec,lens}$\tablenotemark{c}} & \colhead{$z_\text{phot,lens}$\tablenotemark{d}} & \colhead{$z_\text{phot,source}$} & \colhead{Lens model} & \colhead{$\mu$} & \colhead{$m_I$\tablenotemark{e}} & \colhead{$m_{F150W}$\tablenotemark{f}}\\ 
\colhead{} & \colhead{(deg)} & \colhead{(deg)} & \colhead{} & \colhead{} & \colhead{} & \colhead{} & \colhead{} & \colhead{} & \colhead{}} 

%% All data must appear between the \startdata and \enddata commands
\startdata
COSMOS-001\tablenotemark{b} & 150.10031 & 1.89298 & - & 1.17 & $3.1^{+3.1}_{-1.5}$ & SIE+shear & $8.1^{+0.1}_{-0.0}$ & 23.74 & 25.23$\pm$0.03 \\
COSMOS-002 & 150.19857 & 2.11697 & 0.6039 & 0.17 & $6.6^{+0.4}_{-4.4}$ & SIE+shear & $1.7^{+0.0}_{-0.1}$ & 22.17 & 24.15$\pm$0.03 \\
COSMOS-003 & 149.91412 & 2.54399 & 0.6778 & 0.64 & $3.5^{+3.0}_{-1.5}$ & SIE+shear & $1.9^{+0.0}_{-0.0}$ & 21.64 & 26.87$\pm$0.15 \\
COSMOS-004 & 150.02852 & 2.55205 & - & 0.78 & $6.5^{+0.3}_{-4.2}$ & EPL+shear & $3.6^{+0.4}_{-1.4}$ & 19.99 & 23.64$\pm$0.02 \\
COSMOS-005 & 150.05630 & 2.57340 & 0.909 & 1.03 & $4.8^{+0.9}_{-3.2}$ & SIE & $2.4^{+0.0}_{-0.0}$ & 21.54 & 28.08$\pm$0.25 \\
COSMOS-006 & 149.82169 & 1.90706 & 1.3685 & 1.35 & $2.8^{+3.6}_{-0.9}$ & SIE & $4.2^{+0.0}_{-0.0}$ & 23.55 & 23.85$\pm$0.03 \\
COSMOS-007 & 150.44997 & 2.14277 & 0.8891 & 1.05 & $1.3^{+2.5}_{-0.3}$ & EPL & $1.5^{+0.0}_{-0.0}$ & 21.55 & 22.04$\pm$0.01 \\
COSMOS-008 & 149.88085 & 2.12090 & 0.6802 & 0.61 & $3.2^{+2.4}_{-2.1}$ & SIE & $1.4^{+0.0}_{-0.0}$ & 21.48 & 25.31$\pm$0.03 \\
COSMOS-009 & 150.27615 & 2.44736 & 0.22 & 0.22 & $3.0^{+3.1}_{-1.6}$ & EPL & $2.6^{+0.2}_{-0.2}$ & 20.03 & 24.70$\pm$0.03 \\
COSMOS-010 & 150.22805 & 2.55103 & 0.7331 & 0.69 & $3.1^{+0.4}_{-2.0}$ & SIE & $2.3^{+0.3}_{-0.4}$ & 20.97 & 23.63$\pm$0.02 \\
COSMOS-011 & 150.26621 & 2.04930 & 0.96 & 0.99 & $2.6^{+4.1}_{-0.5}$ & SIE+shear & $1.9^{+0.0}_{-0.0}$ & 21.19 & 23.51$\pm$0.02 \\
STUDIES-COS-019\tablenotemark{b} & 150.09994 & 2.29731 & 0.3598 & 0.37 & $2.7^{+0.2}_{-0.7}$ & SIE & $5.4^{+0.0}_{-0.1}$ & 20.13 & 23.30$\pm$0.00 \\
STUDIES-COS-035 & 150.20925 & 2.35513 & 0.1656 & 0.16 & $2.5^{+0.2}_{-0.5}$ & SIE & $2.5^{+0.0}_{-0.0}$ & 18.21 & 21.87$\pm$0.00 \\
STUDIES-COS-049 & 150.16443 & 2.40885 & 0.991 & - & $2.9^{+3.5}_{-1.0}$ & SIE & $2.4^{+0.0}_{-0.0}$ & 22.85 & 23.37$\pm$0.01 \\
STUDIES-COS-174 & 150.11834 & 2.29226 & 1.0995 & 1.15 & $1.6^{+1.6}_{-0.5}$ & SIE & $2.4^{+0.0}_{-0.0}$ & 22.06 & 21.97$\pm$0.00 \\
STUDIES-COS-195 & 150.09114 & 2.45861 & 0.5976 & - & $1.3^{+1.8}_{-0.3}$ & SIE & $2.2^{+0.0}_{-0.0}$ & 21.12 & 22.35$\pm$0.00 \\
STUDIES-COS-218 & 150.05421 & 2.33949 & 0.3778 & 0.37 & $2.9^{+0.1}_{-1.5}$ & SIE+shear & $4.7^{+0.0}_{-0.0}$ & 19.44 & 22.23$\pm$0.00 \\
STUDIES-COS-232 & 150.11882 & 2.32435 & 0.6043 & 0.49 & $1.1^{+0.1}_{-0.1}$ & SIE & $2.7^{+0.0}_{-0.0}$ & 20.45 & 21.76$\pm$0.00 \\
\enddata
\tablenotetext{a}{The IDs STUDIES-COS-XXX are shortened forms of STUDIES-COSMOS-450-XXX, as used in \citet{2024ApJ...971..117G}.}
\tablenotetext{b}{COSMOS-001 is the COSMOS-Web ring discovered by \citet{2024NatAs...8..119V} and \citet{2024Mercier}. STUDIES-COSMOS-450-019 was identified by \citet{2024MNRAS.52712044P}.}
\tablenotetext{c}{Spectroscopic redshifts of the foreground lenses, obtained from hCOSMOS \citep{2018ApJS..234...21D}, zCOSMOS \citep{2007ApJS..172...70L}, zCOSMOS 20k \citep{2014MNRAS.438..717K}, DESI \citep{2024AJ....168...58D}; DEIMOS \citep{2018ApJ...858...77H}, COSMOS-XS \citep{2016MNRAS.463..413W, 2017ApJS..233...19C, 2021ApJ...907....5V, 2023AA_673A_67G}}
\tablenotetext{d}{Photometric redshifts of the foreground lenses estimated from EAZY. The uncertainties are less than 1\%}
\tablenotetext{e}{The I band AB magnitude of the foreground lenses, obtained from COSMOS2020 catalog \citep{2022ApJS..258...11W}}
\tablenotetext{f}{The F150W AB magnitude of the background sources in our threshold catalog (Section~\ref{sec:infrared}).}
\label{table1}
\end{deluxetable*}

\subsection{Submillimeter Data}

The submillimeter data used in this work include observations in COSMOS and UDS fields from the Atacama Large Millimeter Array (ALMA) and the Submillimeter Common-User Bolometer Array 2 (SCUBA-2) mounted on the James Clerk Maxwell Telescope (JCMT). The ultradeep 450\,$\micron$ image from SCUBA-2 was based on the SCUBA-2 Ultra Deep Imaging EAO Survey \citep[STUDIES;][]{2017ApJ...850...37W}, where the latest release \citep{2024ApJ...971..117G} has combined data from other surveys such as the SCUBA-2 Cosmology Legacy Survey \citep[S2CLS;][]{2013MNRAS.432...53G}, the SCUBA-2 COSMOS survey \citep[S2COSMOS;][]{2019ApJ...880...43S}, as well as smaller-scale surveys (e.g., \citealt{2013MNRAS.436.1919C}). The final STUDIES images have reached a noise level of $\sigma_{450}$=0.6\,mJy\,beam$^{-1}$, detecting over 400 DSFGs that are up to one order of magnitude fainter than those detected by Herschel at 500\,$\micron$, making STUDIES an ideal venue to search for strongly lensed faint DSFGs. The JWST counterparts of STUDIES 450\,$\micron$ sources are identified by cross-matching our JWST catalogs with the positions reported in the existing ALMA (AS2COSMOS, A3COSMOS), VLA (1.4 and 3\,GHz), or MIPS catalogs (Gao et al. in preparation). 

In addition to the 450\,$\micron$ data, we also make use of the 850\,$\micron$ image from SCUBA-2, which was sourced from S2COSMOS \citep{2019ApJ...880...43S}. The ALMA imaging data were obtained from the ALMA Science Archive.
% , with flux densities calibrated to mJy/beam. ALMA imaging data were obtained from the ALMA Science Archive, covering multiple projects focused on different galaxies.

\section{Analyses and Results} \label{sec:analyses}
\subsection{Identification of strongly lensed DSFGs}

\begin{figure*}[t!]
    \centering
    \includegraphics[width=.8\linewidth]{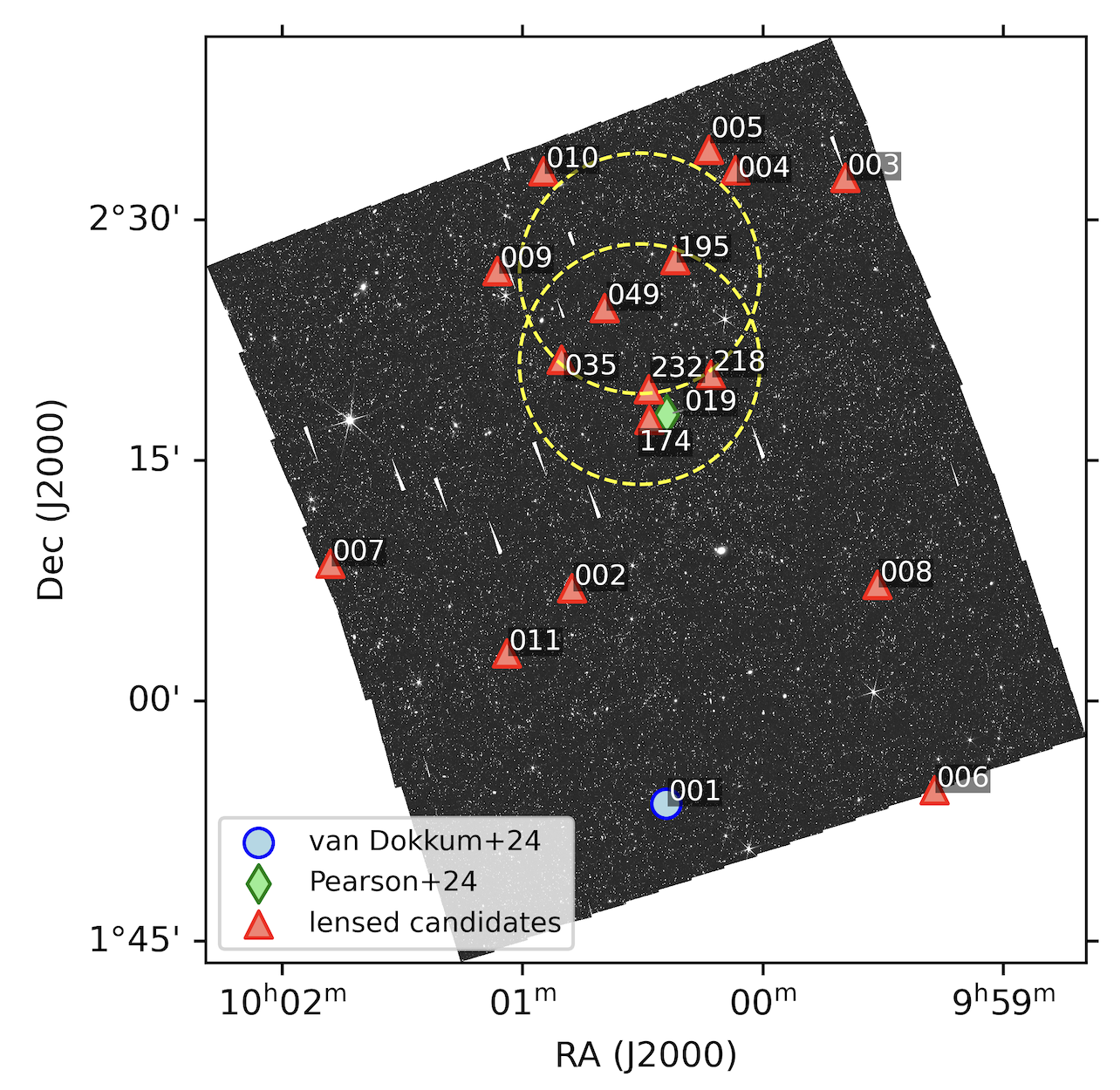}
    \caption{The sky positions of all the 18 strongly lensed faint DSFG candidates identified by this work. The two yellow dashed circles show the footprint of STUDIES. The blue dot and green diamond represent the strongly lensed DSFGs found in previous studies \citep{2024MNRAS.52712044P,2024NatAs...8..119V}, which are also identified by our method.\label{fig:position}}
    % The red triangles represent the DSFG candidates with strong lensing features in JWST images.
    
\end{figure*}
Our method of identifying strongly lensed systems is the following; We first select strongly lensed candidates by visually inspecting the JWST r-g-b (F444W-F277W-F150W) false-color cutout images. Sources that have morphology resembling lensing arcs or rings, or configurations that are consistent with red emission located near foreground large bluer sources, are all selected. At this stage, we take a conservative approach that aims to select all sources that could potentially be lensed. These candidates are then passed to the second stage where we perform redshift analyses and lens modeling in order to confirm their strongly lensed nature. 

The above method is applied to two sets of sources. The first is the STUDIES 450\,$\micron$ catalog, and seven candidates have been identified. Their associations with STUDIES sources are all confirmed via counterpart identifications of ALMA, VLA, or MIPS catalogs. 

% We started the search for strongly lensed STUDIES sources by visually examining their r-g-b (F444W-F277W-F150W) false-color cutout images. Potential lensing arcs or morphological configurations that are consistent with red emission located near foreground large bluer sources.  Seven candidates have been identified this way, and their associations with STUDIES sources are all confirmed via counterpart identifications of ALMA, VLA, or MIPS catalogs.

% We selected our sample based on flux and color selection indicative of dusty galaxies from JWST data. 
To allow the search of faint DSFG across the full COSMOS-Web field, including the regions outside the STUDIES footprint, the second set of sources are identified via a color selection method, where the F444W flux density is required to exceed 1 $\mu$Jy and a flux ratio between F444W and F150W is greater than 3.5 \citep{2023ApJ...956...95B}. The efficacy of these color and flux criteria for the selection of faint DSFGs is further verified recently by \citet{McKay:2025aa}. To consider the color of the extended sources, we used the isophotal flux from the threshold catalog, which is measured in the same segmentation across different bands. The flux in F444W was taken from the original catalog, which were corrected to total. The cut of 10\,$\sigma$ in the threshold catalog corresponds to the F444W-to-F150W flux ratio of about 1, meaning that the process of making the threshold catalog does not lead to missing red sources that satisfy the color criteria. Using these flux and color selection criteria, we identified 11,873 candidate dusty galaxies in the COSMOS-Web field and 1,425 in the PRIMER-UDS field. We visually inspect them to identify candidate strong lensing systems, following the same approach as we did for the STUDIES sources.

To confirm the dusty nature of these red sources, we performed a stacking analysis of submillimeter data. The weighted mean stacking shows a flux density of 0.79$\pm$0.04 mJy/beam at 850\,$\micron$, and 3.08$\pm$0.16 mJy/beam at 450\,$\micron$ in the COSMOS-Web field, indicating submillimeter detections with high signal-to-noise ratios. The stacking in the PRIMER-UDS field has yielded consistent results, and stacking instead in the median also does not change the results significantly. Eventually, in this second set of red sources, we identify 13 strongly lensed candidates, again all in the COSMOS field.
% In the PRIMER-UDS field, the weighted stacked flux density is 0.73$\pm$0.04 mJy/beam in 850-micron image, similar to the result of COSMOS field.
% Furthermore, to include more dusty galaxies which may be not red enough to meet the criteria, we incorporated the counterparts of STUDIES 450-micron as additional sample groups.

% To search for strongly gravitationally lensed systems, we generated red-green-blue (RGB) composite images from NIRCam image in the F444W, F277W, and F150W filters and created cutout images for each candidate galaxy. Strongly lensed sources were identified by visually inspecting these cutout images for arc or ring structures near another galaxy, which can potentially be the foreground lens. 

Therefore, in total, we identified 18 potential strongly lensed sources in the COSMOS field. Of these, 13 were selected from the red JWST sources, and 7 were identified from the STUDIES 450\,$\micron$ catalog, with two of the sources overlapped. The source coordinates are given in Table~\ref{table1}, and the overview of their locations on the COSMOS-Web footprint is shown in Figure~\ref{fig:position}. Note that none of the candidates are found in the PRIMER-UDS field, which could be attributed to the 10 times smaller footprint and possibly field-to-field variations. 

Among the selected sources, COSMOS-001 and STUDIES-COSMOS-450-019 \citep[COSBO-7 in][]{2024A&A...690L..16J} have been found and confirmed as strongly lensed in previous studies \citep{2024MNRAS.52712044P, 2024NatAs...8..119V, 2024Mercier}. The latter source had also been identified as a potential lensing candidate in earlier studies \citep{2018ApJ...864...56J,2021ApJ...913....6H}. \citet{2018ApJ...864...56J} also mentioned the other two lensing candidates, one of which is found in our work as COSMOS-005 (ID85004261, AzTEC/C10), but the other lies outside the COSMOS-Web area.

\begin{figure*}[t!]
    \centering
    \includegraphics[width=\linewidth]{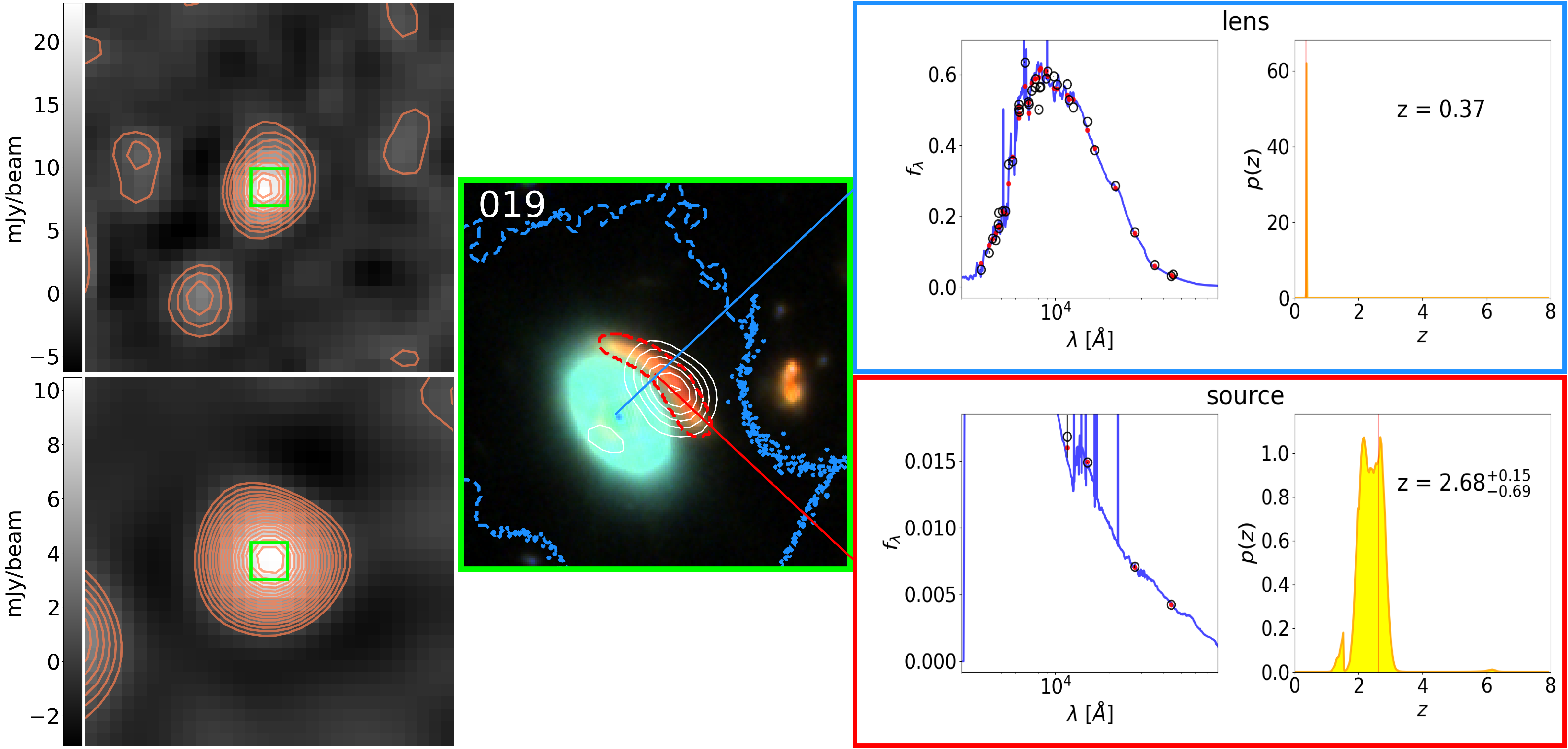}
    \caption{The cutout image and the photo-$z$ analysis of STUDIES-COSMOS-450-019. The left upper and lower panels show 1$'$ cutout images of the SCUBA-2 450 $\mu$m and 850 $\mu$m maps, respectively, with orange contours starting at 1$\sigma$ with a step of 1$\sigma$. The green squares mark the position and size of the JWST cutout image shown in the middle panel. The middle panel presents a 6$''$ RGB (F444W-F277W-F150W) composite image generated from NIRCam data, with white ALMA contours, starting at 3$\sigma$ with a step of 1$\sigma$.
    % with white solid and dashed contours from VLA at 3 GHz and MIPS at 24 $\mu$m, respectively. The VLA contours starts at 3$\sigma$ with a step of 2$\sigma$.
    % where F150W, F277W, and F444W are assigned to the blue, green, and red channels, respectively. White ALMA contours, starting at 3$\sigma$, are overlaid to highlight regions of significant submillimeter emission. 
    The right panels display the results of the photo-$z$ analysis, including SED fitting and the redshift probability distributions. Vertical lines mark the spectroscopic redshifts obtained from previous studies. The aperture used for flux measurement is indicated in the middle panel.
    \label{fig:cutout_photz}}
\end{figure*}

For these 18 lensed candidates, we checked their individual submillimeter detections using SCUBA-2 and ALMA data. Seven of the 18 show individual detections over 3$\sigma$ in the SCUBA-2 850$\micron$ data, and 450\,$\micron$ detections are only found from the STUDIES sources. Looking at ALMA archival data, we found that 15 sources had been observed and three of them exhibit detections above 3$\sigma$. The rest of the non-detection in ALMA can be attributed to too low of sensitivity. An example source with SCUBA-2 450\,$\micron$ and JWST cutout images is shown in Figure~\ref{fig:cutout_photz}, and the rest of the source cutouts are shown in the Appendix.

% However, no potential strongly lensed DSFG were found in the candidate dusty galaxies selected with the flux and color criteria in PRIMER-UDS field, exhibiting a variance of the number of the source in different field. We also made 10”$\times$10” cutouts of STUDIES 450-micron sources to look if there are potential lensing system, but no obvious strong lensing feature were found.

\subsection{Photometric redshift}

For the strongly gravitational lensed sources and their foreground lenses, we obtained photometric redshifts (photo-$z$) using EAZY \citep{2008ApJ...686.1503B}. The spectral energy distribution (SED) of the lensed sources was fitted using the EAZY template set from \citet{2008ApJ...686.1503B} that accounts for dusty galaxies, with photometry from the four JWST bands (F115W, F150W, F277W, and F444W). For the foreground galaxies, SED fitting was performed with also EAZY template in 36 to 38 bands, including the four JWST bands and additional photometry listed in the COSMOS2020 catalog \citep{2022ApJS..258...11W}. An example set of SED fitting results and the probability distributions of photometric redshifts for one system is shown in Figure~\ref{fig:cutout_photz}, and the redshifts are given in Table~\ref{table1}.

In our 18 potential lensing systems, 16 lenses have spectroscopic redshifts, which are obtained from past studies \citep{2024AJ....168...58D, 2023AA_673A_67G, 2021ApJ...907....5V,  2018ApJ...858...77H, 2018ApJS..234...21D, 2017ApJS..233...19C, 2016MNRAS.463..413W, 2014MNRAS.438..717K, 2007ApJS..172...70L}. Our photo-$z$ results are consistent with spectroscopic redshifts for most lenses. Two lenses do not have good photo-$z$ solutions, however they do have spectroscopic redshifts.
% two lenses seem to be outliers in the photo-$z$ analysis, as no good fitting results were obtained even after trying different templates or removing some bands. There are also no reliable photo-$z$ estimates for these two lenses in the COSMOS2020 catalog. However, using the spec-z data for these lenses, we could still focus on the foreground and background relation between the lenses and sources in these two systems and evaluated whether they can potentially be a lensing system.

Recently, a study estimated the photometric redshift of 019 (COSBO-7) to be greater than 7 by fitting 10 NIRCam bands using four independent SED fitting codes \citep{2024ApJ...969L..28L}. However, this estimate was soon refuted by the spectroscopic redshift of z = 2.625, confirmed by robust lines detected by ALMA \citep{2024A&A...690L..16J}. Our photometric redshift estimation for this source aligns with the spectroscopic redshift for both lens and source in the 019 lensing system, again validating our redshift analyses.

Although the photometric redshifts of the background sources have large uncertainties, all lensed sources exhibit higher photo-$z$ values compared to their foreground lenses, confirming the necessary conditions for gravitational lensing.

\subsection{Lens modeling}

To confirm the strongly lensed nature of the selected systems, which is typically defined as having a lensing magnification greater than 2, we performed detailed foreground lens modeling using Lenstronomy \citep{2021JOSS....6.3283B, 2018PDU....22..189B}, a multi-purpose gravitational lens modeling software package based on \citet{2015ApJ...813..102B}. We first mask the observed 6$''\times6''$ images using the segmentation map to isolate the relevant components, minimizing contamination from unrelated sources.

\begin{figure*}[t!]
    \centering
    \includegraphics[width=\linewidth]{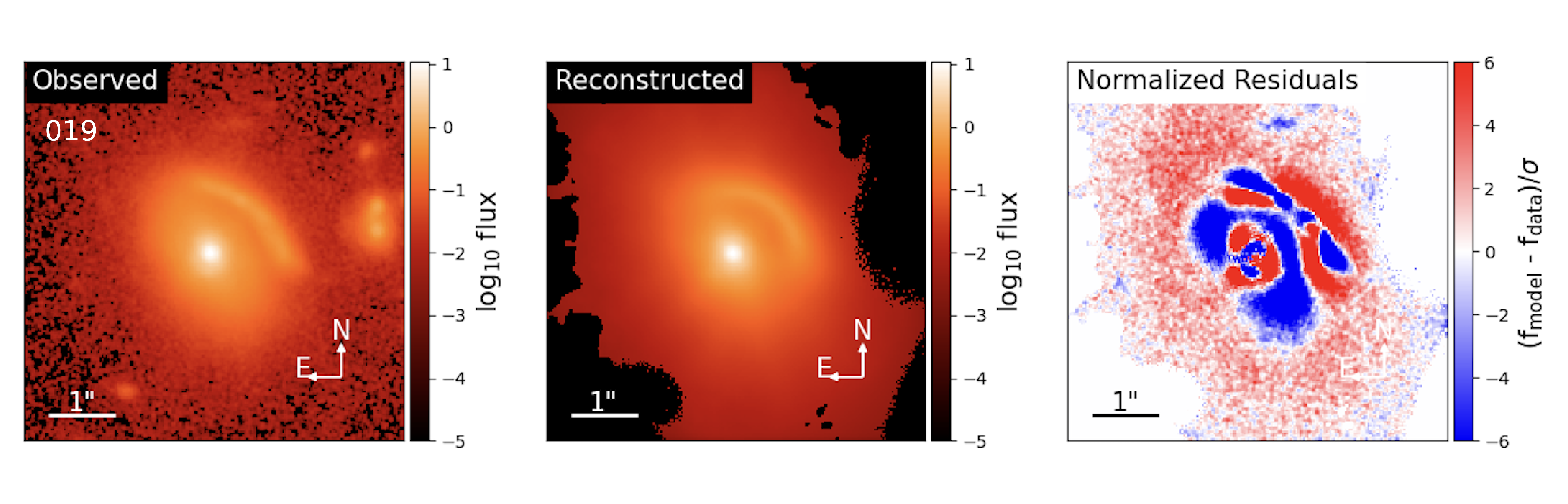}
    \caption{An example of the result from lens modeling for STUDIES-COSMOS-450-019. From left to right: the observed NIRCam F444W image centered on the foreground lens, the reconstructed model image with the applied mask, and the residual between the model and the observed image. All images have the same size, orientation, and display flux in logarithmic scale.
    \label{fig:lens_modeling}}
\end{figure*}

In the modeling process, we input the mass distribution for the foreground lens and the light profiles for both the lens and the source. By constructing a model image, we were able to fit the observed data and obtain the best-fit parameters. For the model of foreground mass distribution, we tried Singular Isothermal Ellipse (SIE) and Elliptical Power Law (EPL), with and without cosmic shear, and selected the best fit model for each system. Sérsic ellipse is considered as the light profile for both lens and source. For COSMOS-005, which has a more complex foreground, we considered 2 Sérsic ellipse to obtain a better fitting result. The fitting was performed simultaneously on images from two bands, F444W and F150W, using Particle Swarm Optimization (PSO) and Markov Chain Monte Carlo (MCMC) methods. An example of the constructed model image and the residual is shown in Figure~\ref{fig:lens_modeling}. The modeling results of the best-fit mass distribution models and magnification factors for each system are exhibited in Table~\ref{table1}. Among the 18 candidates, 13 are confirmed strong lensing systems with magnifications greater than 2. These 13 sources are then passed to the next stage for the construction of number counts.

% comparison with COWLS
Recently, over 100 candidates of strong galaxy-galaxy lensing have been reported in the COSMOS-Web field by \citet{Nightingale:2025aa}, and we find that nine of our sources can be matched to theirs, where five are matched to STUDIES sources and four to the color-selected sources. Since both our work and that of \citet{Nightingale:2025aa} rely on visual inspection, it is possible that the mismatch is caused by some level of incompleteness.
% After inspecting their method we conclude that the most probable reason for the mismatch for the rest of the nine sources is because \citet{Nightingale:2025aa} imposed a magnitude cut for selecting bright and massive foreground lens, whereas we do not impose such a cut. Sources such as COSMOS-002 and STUDIES-COSMOS-450-049 (Figure~\ref{fig:A1}, \ref{fig:A2}) are systems where the foreground lenses are not particularly bright. 
For the nine matches, we find a good agreement in the reported lensing magnification, with a median ratio of 1.2$\pm$0.3, where the values reported by \citet{Nightingale:2025aa} are slightly higher. This confirms our lens modeling method and the strong lensing nature of our sample. 

\subsection{Number counts of strongly lensed faint DSFGs}

% We divided our sample into two groups based on their submillimeter flux: (1) the 450\,$\micron$ counterparts of STUDIES sources and (2) the submillimeter-faint strongly lensed DSFGs identified with JWST flux and color selection. The cumulative number counts of the strongly lensed sources are plotted in two groups and compared with the model predictions in Figure~\ref{fig:density-flux}. 

The cumulative number counts of strongly lensed STUDIES sources are calculated following the well-established methods for constructing counts outlined in the original SCUBA-2 survey \citep{2024ApJ...971..117G}, where corrections for spurious fraction, flux boosting, and completeness are properly taken into account via extensive simulations. We note that the total area considered is the combined area of STUDIES footprint in both COSMOS and UDS. The results are given in Table 2 in the Appendix and plotted in Figure~\ref{fig:density-flux}, together with the counts in the literature and model prediction. Our measurements follow the extrapolation of the bright counts and extend to a flux limit about seven times fainter than the previous limit. Our counts are slightly higher than the pilot search within the STUDIES footprint \citep{2024MNRAS.52712044P} due to more sources discovered. Our results are higher than the model predictions from \citet{Sedgwick:2025aa} but consistent with those from \citet{2017MNRAS.465.3558N}, which appears to be successful in reproducing the observed counts, both lensed and unlensed, across two orders of magnitude in fluxes. The different predictions from the two models highlight the need to conduct more extensive search of faint lensed DSFGs in order to perform more rigorous tests.

It is worth mentioning that while the JWST counterparts of the STUDIES sources are confirmed via higher resolution images from ALMA, VLA, and MIPS, two systems that do not have ALMA data, STUDIES-COSMOS-450-035 and STUDIES-COSMOS-450-218, appear to have their MIPS or VLA emissions located closer to the foreground lens. The blue nature of the lenses suggests that they are not likely to be the origin of the  450\,$\micron$ emissions. What is more likely is that the background lensed dusty sources have a second lensed image that is closer to the foreground lens. This can be seen in STUDIES-COSMOS-450-019 and STUDIES-COSMOS-450-232. However, in order to definitely rule out the possibility of 450\,$\micron$ emissions coming from the foreground lenses, follow-up observations are needed.

\begin{figure*}[t!]
    \centering
    \includegraphics[width=.8\linewidth]{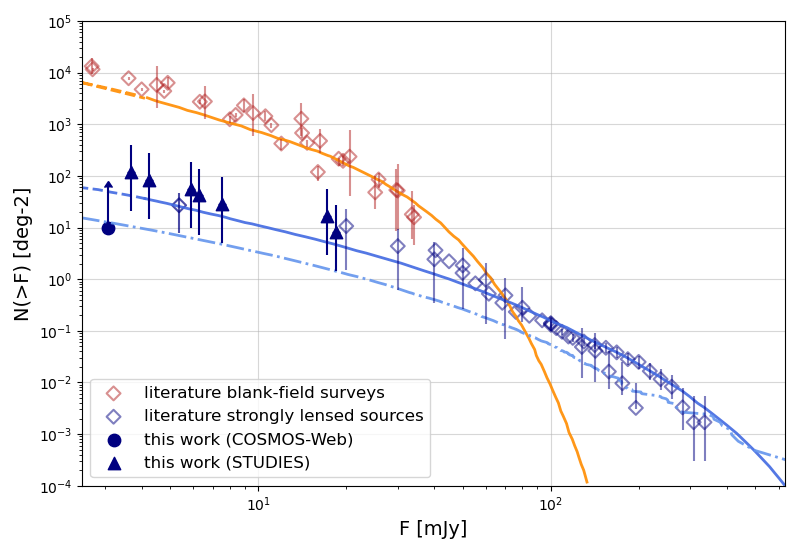}
    \caption{Galaxy cumulative number counts at 450–500\,$\mu$m. The orange curve represent the unlensed model obtained from \citet{Cai2013}. The lensed model from \citet{2017MNRAS.465.3558N} and \citet{Sedgwick:2025aa} are shown in blue solid and dashdot line, respectively. The blue filled data points represent the results of this work. The STUDIES sample results, derived from the effective area and corrected for spurious detections and completeness, are shown as blue triangles. For the flux- and color-selected DSFGs, we derived the stacked submillimeter flux and present the results as a lower limit due to the likely incompleteness, shown as blue dot. The strongly lensed DSFGs identified in \citet{2013ApJ...762...59W, 2017MNRAS.465.3558N, 2022MNRAS.510.2261W} and \citet{2024MNRAS.527.8865B} are shown as blue diamonds. Red data points show the number density of unlensed DSFGs from \citet{2013MNRAS.436.1919C, 2013ApJ...776..131C,2017MNRAS.464.3369Z, 2024ApJ...971..117G} and \citet{2024ApJ...964L..32H}.
    \label{fig:density-flux}}
\end{figure*}

% The brightest source may influence the cumulative number density. 
For the red DSFGs identified in the wider COSMOS-Web footprint, we calculated the number density based on the survey area of both COSMOS-Web and PRIMER-UDS and plotted the result in Figure~\ref{fig:density-flux} at the weighted stacked submillimeter flux density. The counts appear lower than the model prediction. However, the red color selection method, while proven efficient, is likely incomplete. In fact, in the latest study by \citet{McKay:2025aa}, they found that the completeness of the color-base method to be about 80\% for DSFGs selected at 870\,$\mu$m or 1.1/1.2\,mm. Among the seven strongly lensed DSFGs identified in STUDIES, only two meet the color selection criteria, confirming incompleteness. We therefore treat the color-base measurement as a lower limit, implying that our results are consistent with the model. 
% as the faintest submillimeter flux density in our sample remained uncertain. 
% Our method of visually searching for arc- or ring-like structures produced for strong lensing is likely incomplete, potentially missing other strongly lensed systems such as group lenses, which has been observed in other studies (e.g., \citep{Chen2022}). 

% As a result, we treat our measured number counts based on the red DSFGs as the lower limit. Our results again agree with the model prediction, suggesting a population of faint strongly lensed DSFGs. 
% limiting our results to a lower bound on the number density of strongly lensed DSFGs. This lower-limit finding constrains the submillimeter-faint end of the theoretical model and confirms the existence of faint strongly lensed DSFGs, consistent with the model predictions.

Lastly, the number density of strongly lensed DSFGs in our results exhibited field-to-field variation. All strongly lensed DSFGs in our sample are located in the COSMOS-Web field, which is approximately 10 times larger than the PRIMER-UDS field. Assuming the same number density as in the COSMOS field, we would expect to find one or two strongly lensed DSFGs in the PRIMER-UDS field; however, none has been found. The difference is more apparent when we consider only the STUDIES footprint, which is comparable between COSMOS and UDS. One possibility is that the STUDIES footprint in the COSMOS field covers overdense foreground. Indeed, the weak lensing analyses have shown that the STUDIES-COSMOS field exhibits enhanced E-mode shear correlation signals, suggesting a region of stronger lensing \citep{Massey:2007aa}.

\section{Summary} \label{sec:sum}

We have conducted a systematic search of strongly lensed submillimeter-faint dusty galaxies using SCUBA-2 and JWST data in the COSMOS and UDS fields. In the COSMOS field, we identified 18 strongly lensed candidates, including seven sources individually detected in STUDIES and 11 JWST flux- and color-selected DSFGs. To confirm their lensing nature, we derived photometric redshifts for both the foreground lenses and background sources and performed foreground lens modeling with high-resolution JWST imaging data. Among the 18 candidates, 13 are confirmed strong lensing with magnification greater than 2. Our findings provide an observational constraint on the number density of strongly lensed faint DSFGs, which is about 1\% relative to the unlensed sources at the same flux range. Comparing our results with theoretical predictions, we find an agreement within $\rm \pm\,1-\sigma$ uncertainty, particularly constraining the faint end of the model. However, no strongly lensed DSFGs were found in the PRIMER-UDS field, suggesting a field-to-field variation.

The discovery of these systems offers a unique opportunity to probe the early universe and study dusty galaxies. The lensing magnification allows us to investigate the physical properties of intrinsically faint galaxies that could otherwise remain undetected. Detailed follow-up observations with JWST NIRSpec or ALMA could provide spectroscopic confirmation of redshifts and further characterize the properties of these galaxies.

The unparalleled imaging capabilities of JWST in both depth and resolution have recently allowed the discovery of over 100 strong galaxy-galaxy lenses \citep{Nightingale:2025aa}. In addition, the Euclid mission has begun to produce candidates for strong galaxy-galaxy lensing \citep{Nagam2025,Euclid:2025aa}, with the prospect of detecting more than 10,000 faint DSFGs in its full survey \citep{Sedgwick:2025aa}. Indeed, considering the depths that Euclid will reach (e.g., $I_{\rm E}$ = 26.3 and $Y_{\rm E}$/$J_{\rm E}$/$H_{\rm E}$ = 24.5; \citealt{Euclid:2022aa}), all foreground lenses and most background lensed DSFGs of our sample are expected to be detected. It is therefore expected that the statistical constraint on strongly lensed DSFGs will improve by orders of magnitude in the coming years.

%% Also note that the akcnowlodgment environment does not support long amounts of text. If you have a lot of people and institutions to acknowledge, do not use this command. Instead, create a new \section{Acknowledgments}.
% \begin{acknowledgments}
\section{Acknowledgments}
T.-K.Y and C.-C.C. and acknowledge support from the National Science and Technology Council of Taiwan (111-2112-M-001-045-MY3), as well as Academia Sinica through the Career Development Award (AS-CDA-112-M02).
We are grateful to the maintenance and administrative staff of our institutions, whose efforts to support our day-to-day work environment make our scientific discoveries possible. This work is based on observations made with the NASA/ESA/CSA James Webb Space Telescope. The data were obtained from the Mikulski Archive for Space Telescopes at the Space Telescope Science Institute, which is operated by the Association of Universities for Research in Astronomy, Inc., under NASA contract NAS 5-03127 for JWST. These observations are associated with program \#1727 and \#1837. The James Clerk Maxwell Telescope is operated by the East Asian Observatory on behalf of The National Astronomical Observatory of Japan; Academia Sinica Institute of Astronomy and Astrophysics; the Korea Astronomy and Space Science Institute; the National Astronomical Research Institute of Thailand; Center for Astronomical Mega-Science (as well as the National Key R\&D Program of China with No. 2017YFA0402700). Additional funding support is provided by the Science and Technology Facilities Council of the United Kingdom and participating universities and organizations in the United Kingdom and Canada. Additional funds for the construction of SCUBA-2 were provided by the Canada Foundation for Innovation. The authors wish to recognize and acknowledge the very significant cultural role and reverence that the summit of Maunakea has always had within the indigenous Hawaiian community. We are most fortunate to have the opportunity to conduct observations from this mountain. This paper makes use of the following ALMA data: ADS/JAO.ALMA 2021.1.00225.S, 2021.1.01556.S, 2018.1.00231.S, 2018.1.01128.S, 2017.1.00893.S, 2016.1.00171.S, 2016.1.00463.S, 2016.1.00478.S, 2016.1.00726.S, 2016.1.00798.S, 2016.1.01208.S, 2015.1.00055.S, 2013.1.00118.S, 2013.1.01292.S. ALMA is a partnership of ESO (representing its member states), NSF (USA) and NINS (Japan), together with NRC (Canada), NSTC and ASIAA (Taiwan), and KASI (Republic of Korea), in cooperation with the Republic of Chile. The Joint ALMA Observatory is operated by ESO, AUI/NRAO and NAOJ.
JP and SS are supported by the ACME, ELSA, and OSCARS projects. "ACME: Astrophysics Centre for Multimessenger studies in Europe", "ELSA: Euclid Legacy Science Advanced analysis tools", and "OSCARS: Open Science Clusters' Action for Research and Society" are funded by the European Union under grant agreement no. 101131928, 101135203, and 101129751, respectively; ELSA is also funded by Innovate UK grant 10093177. Views and opinions expressed are however those of the authors only and do not necessarily reflect those of the European Union.

% \end{acknowledgments}


\vspace{5mm}
\facilities{JWST (NIRCam), JCMT (SCUBA-2), ALMA}


\software{astropy \citep{2013A&A...558A..33A,2018AJ....156..123A}, EAZY \citep{2008ApJ...686.1503B}, Lenstronomy \citep{2021JOSS....6.3283B, 2018PDU....22..189B} 
          % Cloudy \citep{2013RMxAA..49..137F}, 
          Source Extractor \citep{Bertin1996}
          }



\appendix

\begin{figure}[t!]
    \centering
    \begin{tabular}{cc}
    %\centering
        \includegraphics[width=0.49\textwidth]{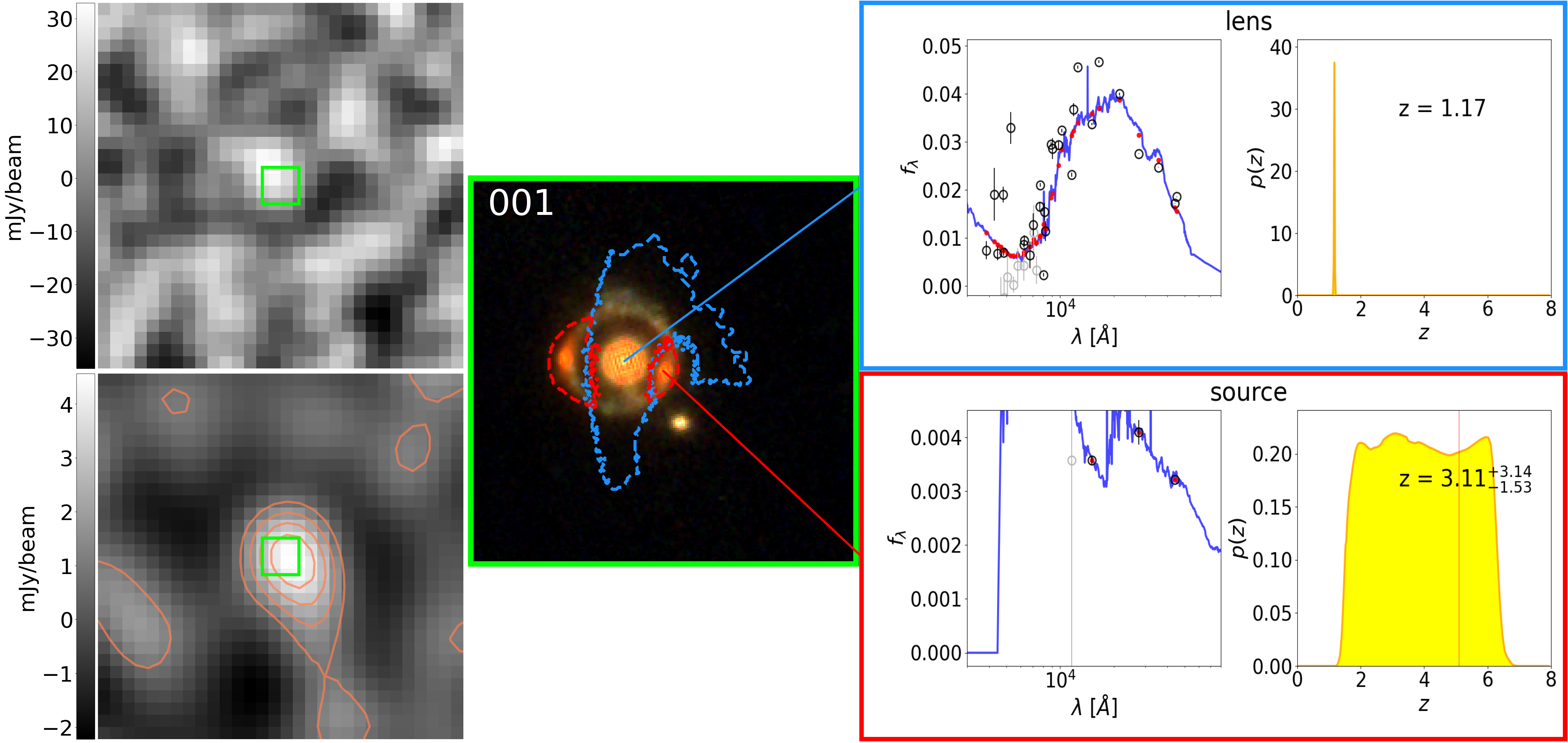} & 
        \includegraphics[width=0.49\textwidth]{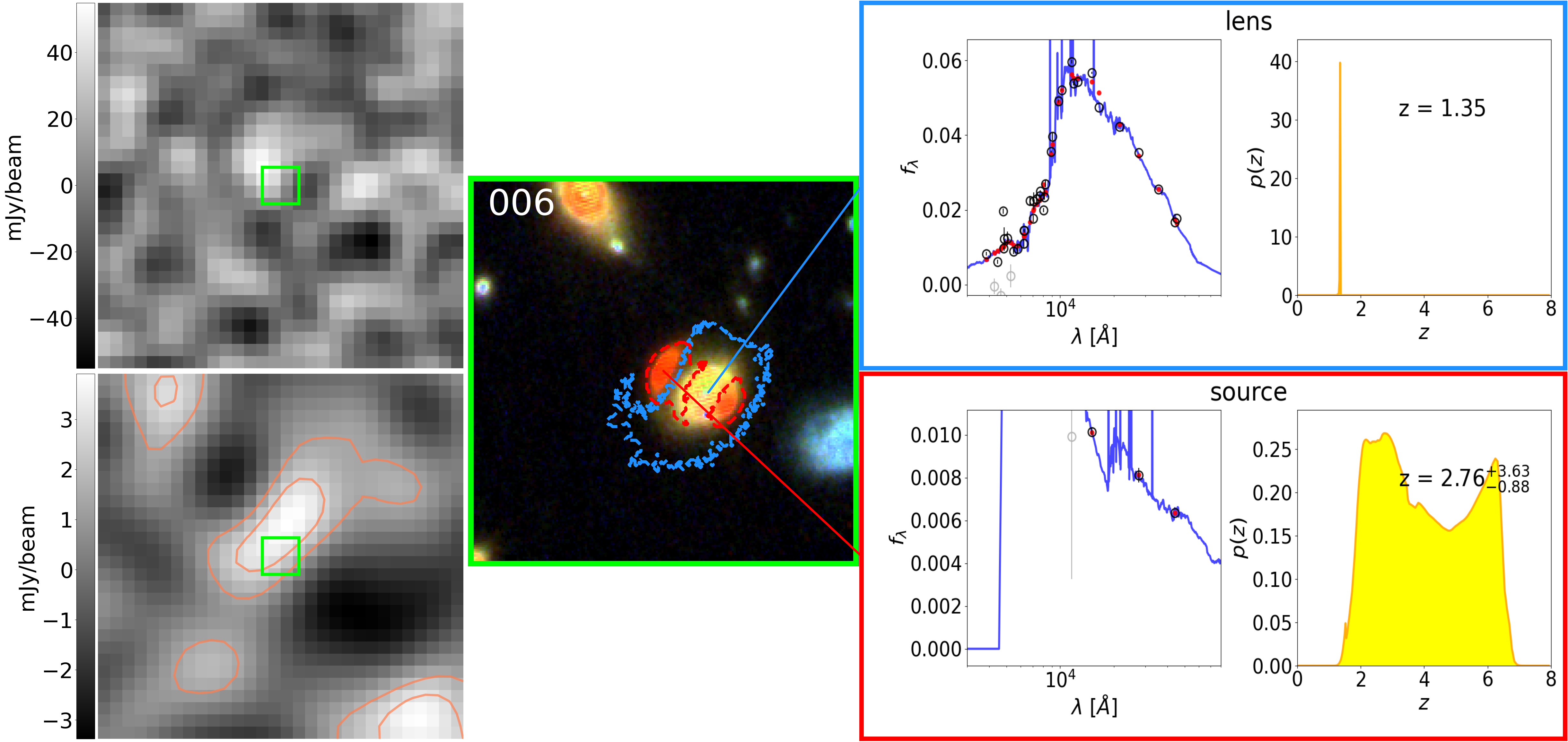} \\     
        \includegraphics[width=0.49\textwidth]{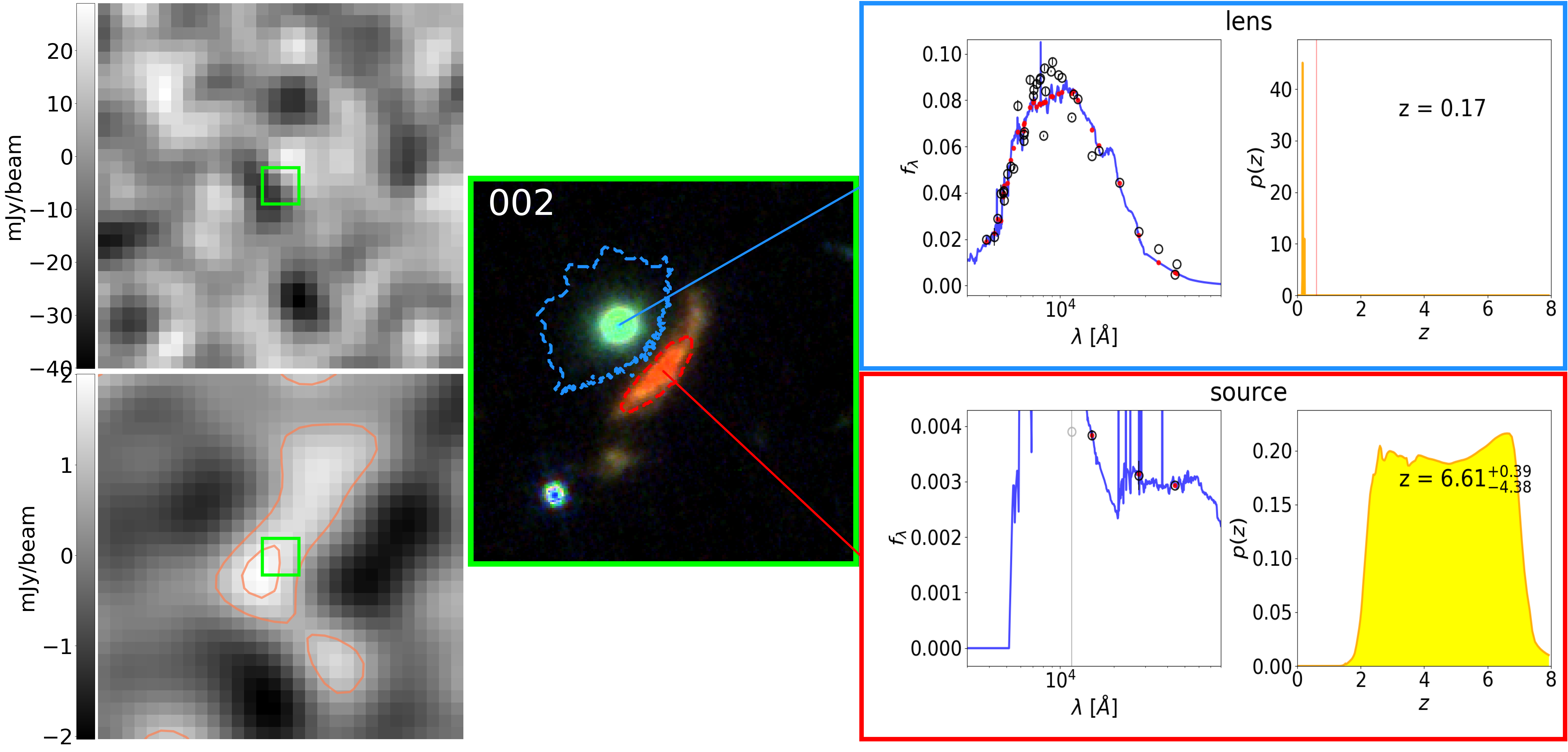} & 
        \includegraphics[width=0.49\textwidth]{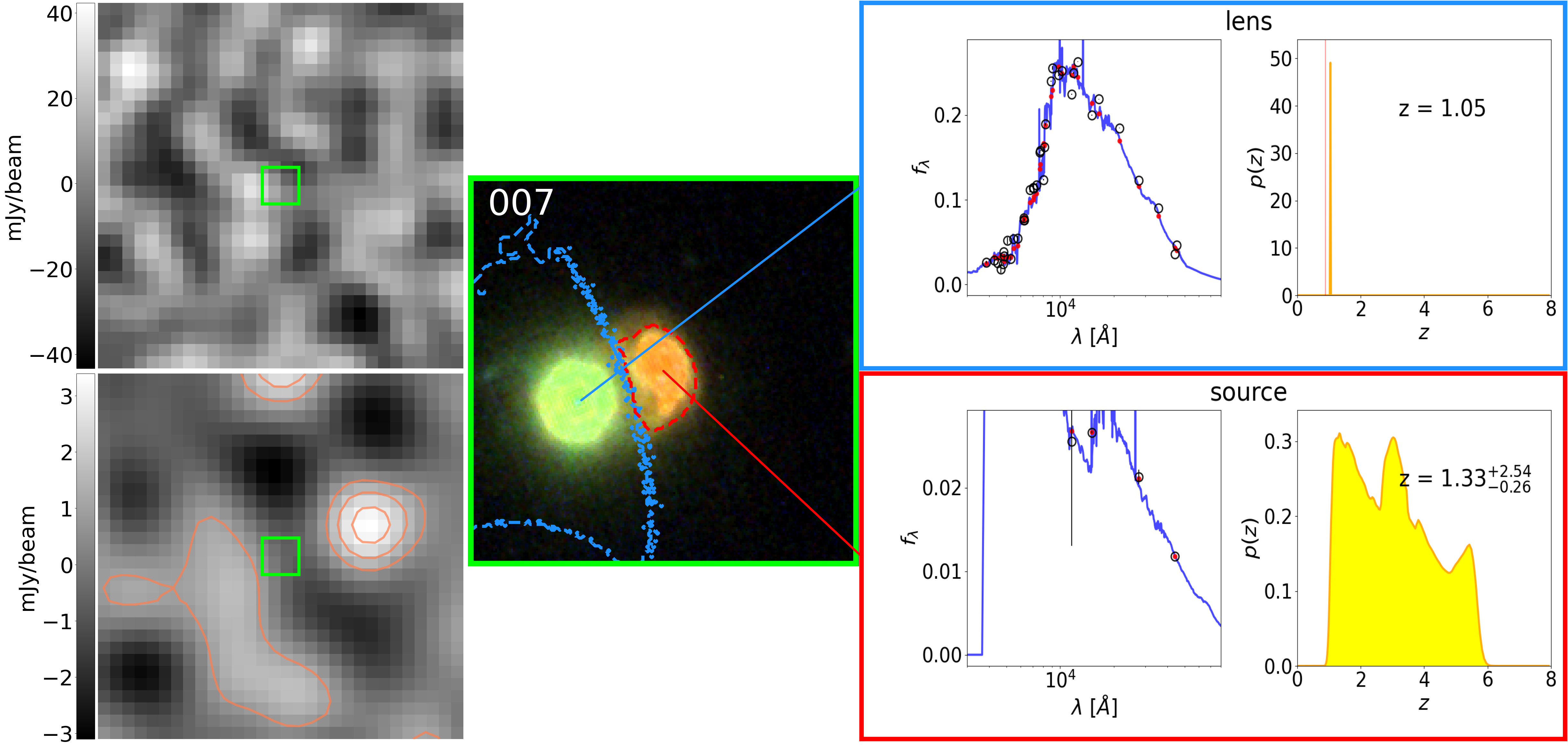} \\ 
        \includegraphics[width=0.49\textwidth]{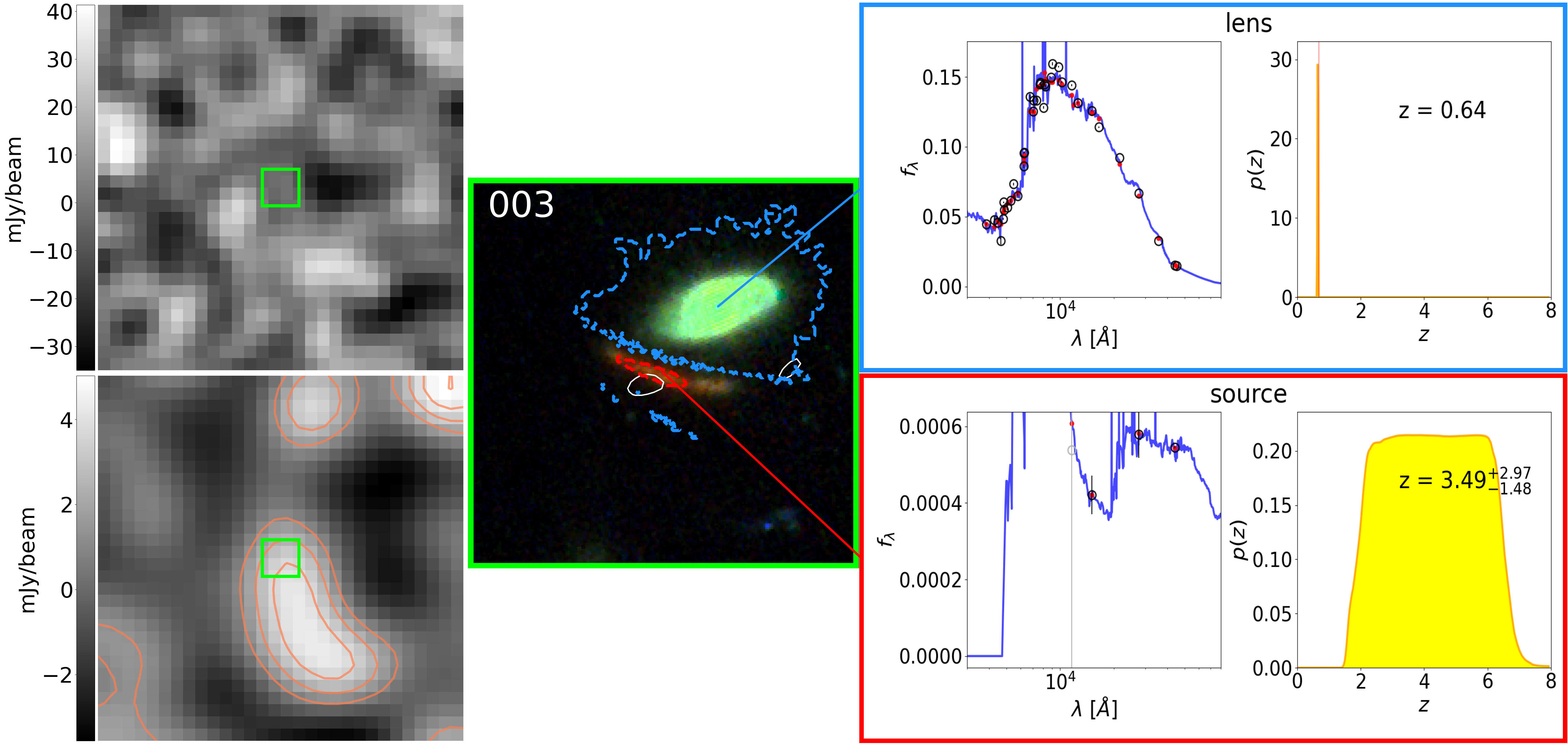} & 
        \includegraphics[width=0.49\textwidth]{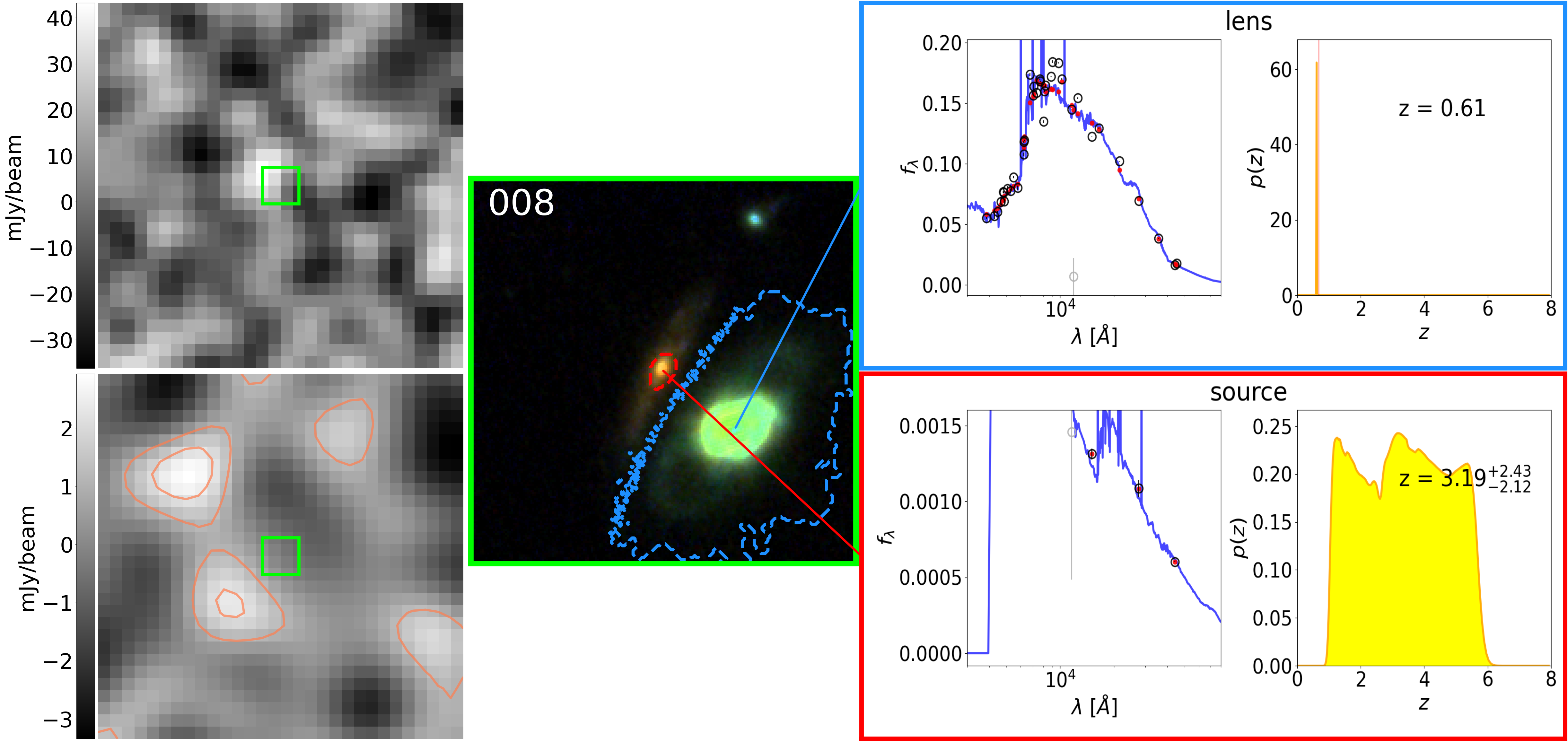} \\ 
        \includegraphics[width=0.49\textwidth]{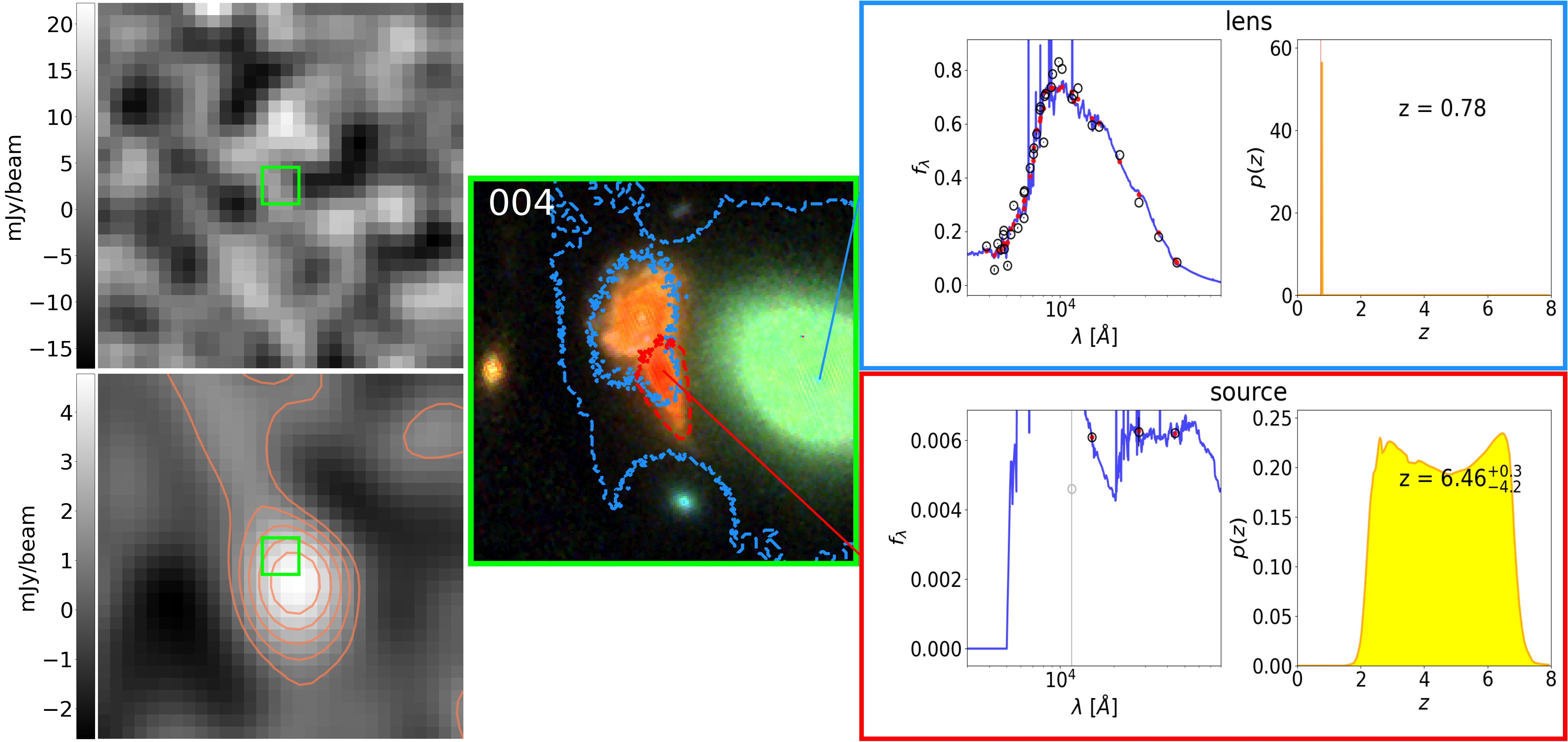} & 
        \includegraphics[width=0.49\textwidth]{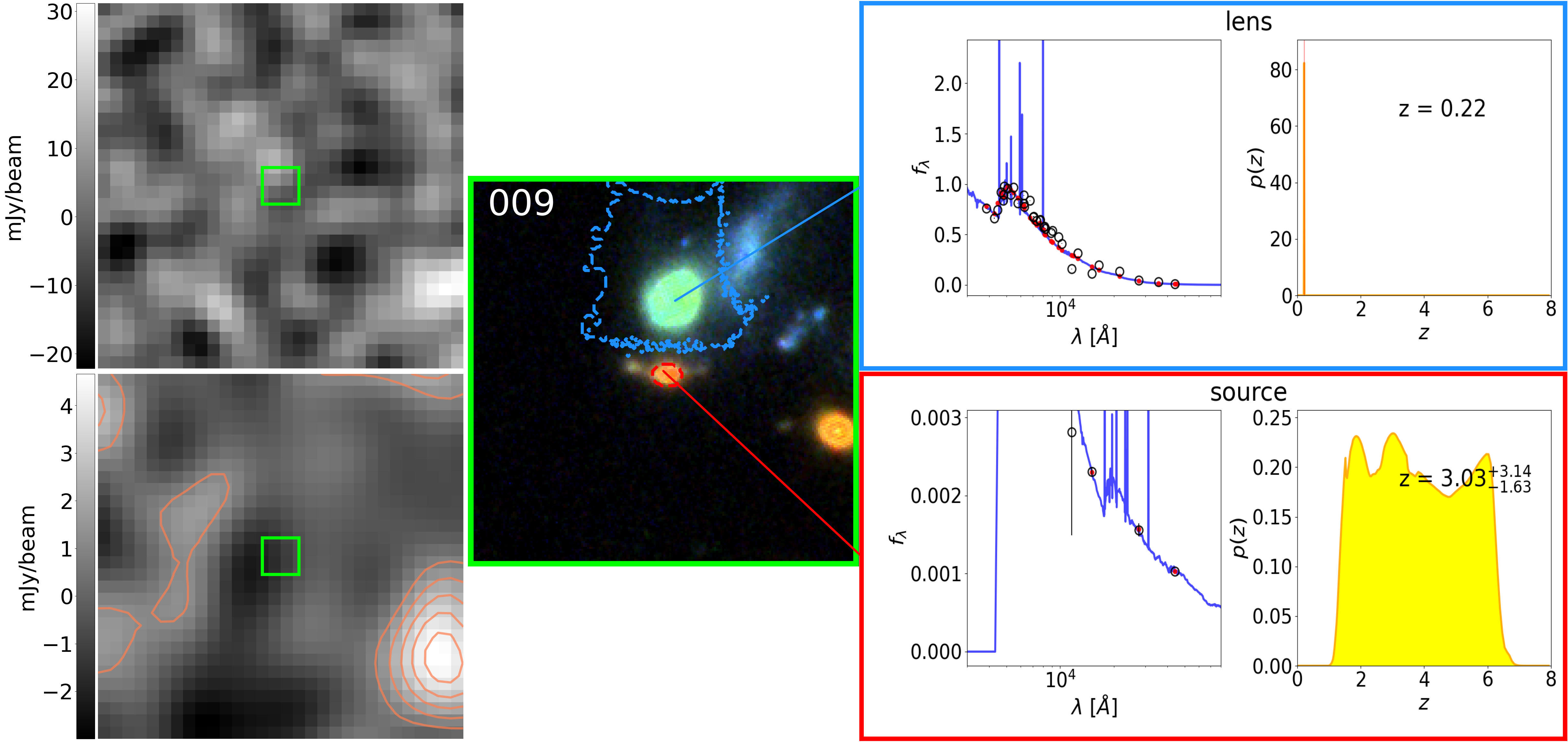} \\ 
        \includegraphics[width=0.49\textwidth]{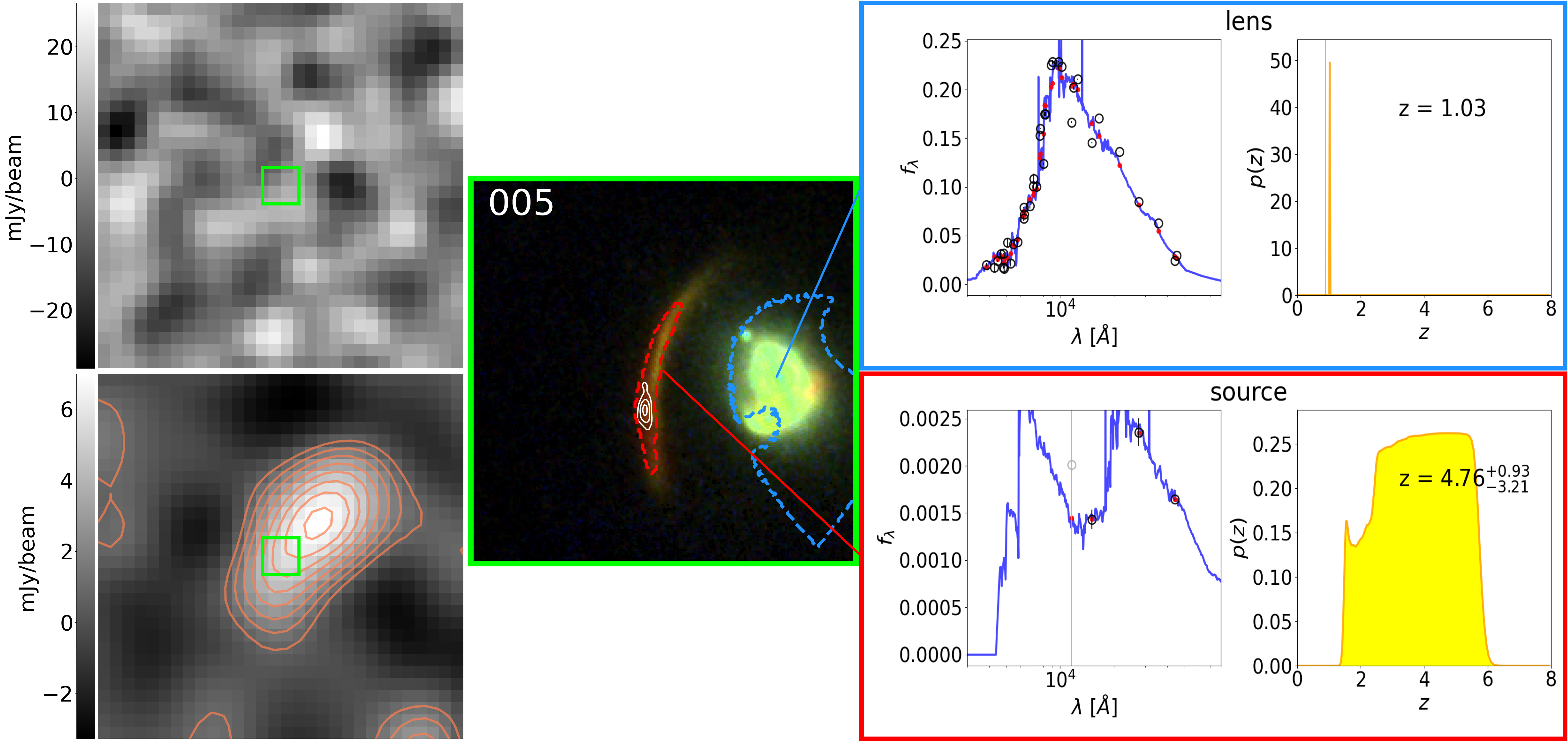} & 
        \includegraphics[width=0.49\textwidth]{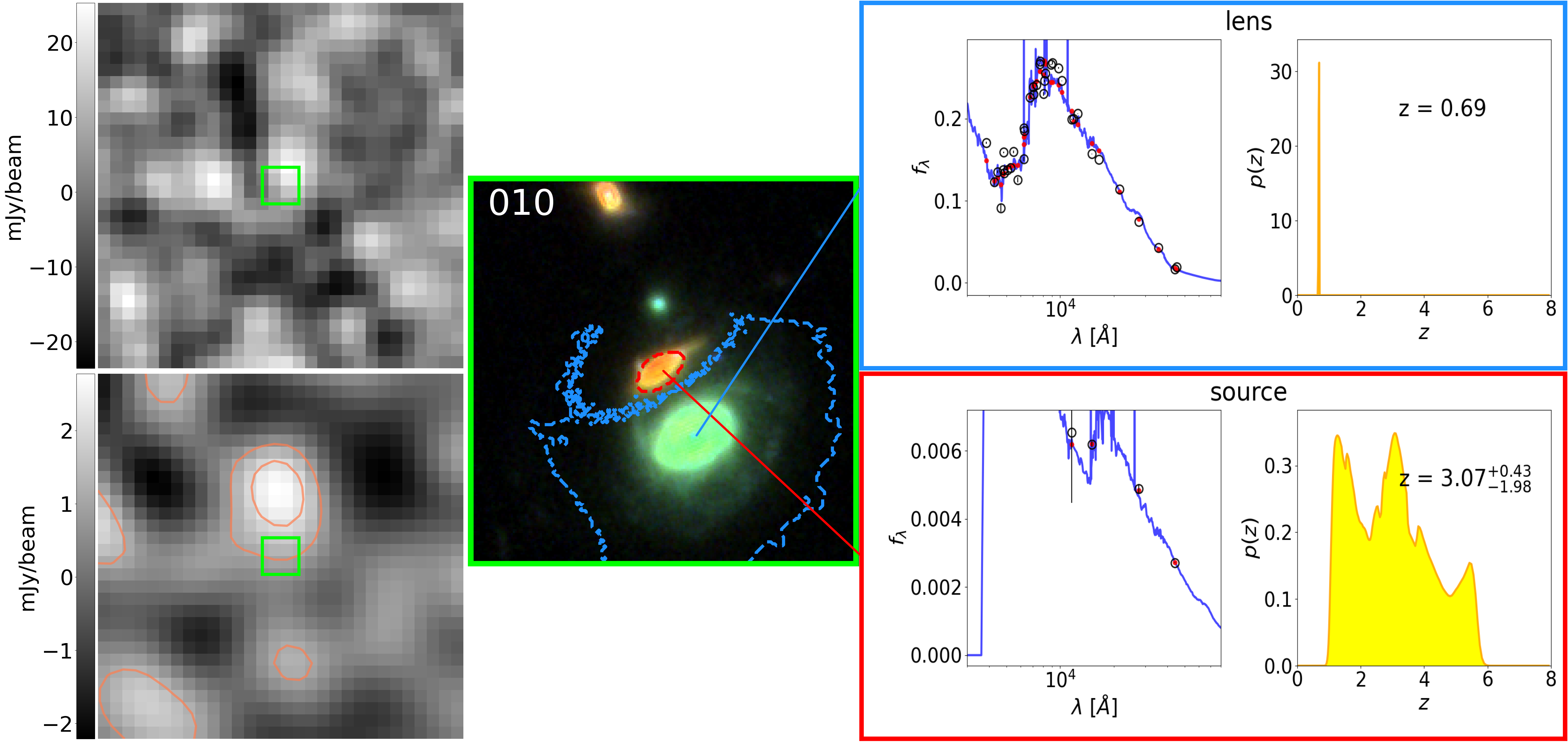} \\ 
        
    \end{tabular}
    \caption{Cutout images and the photo-$z$ analysis for all the 18 candidates, following the style of Figure~\ref{fig:cutout_photz}.}
    \label{fig:A1}
\end{figure}

\begin{figure}[t!]
    \centering
    \begin{tabular}{cc}
    %\centering
        \includegraphics[width=0.49\textwidth]{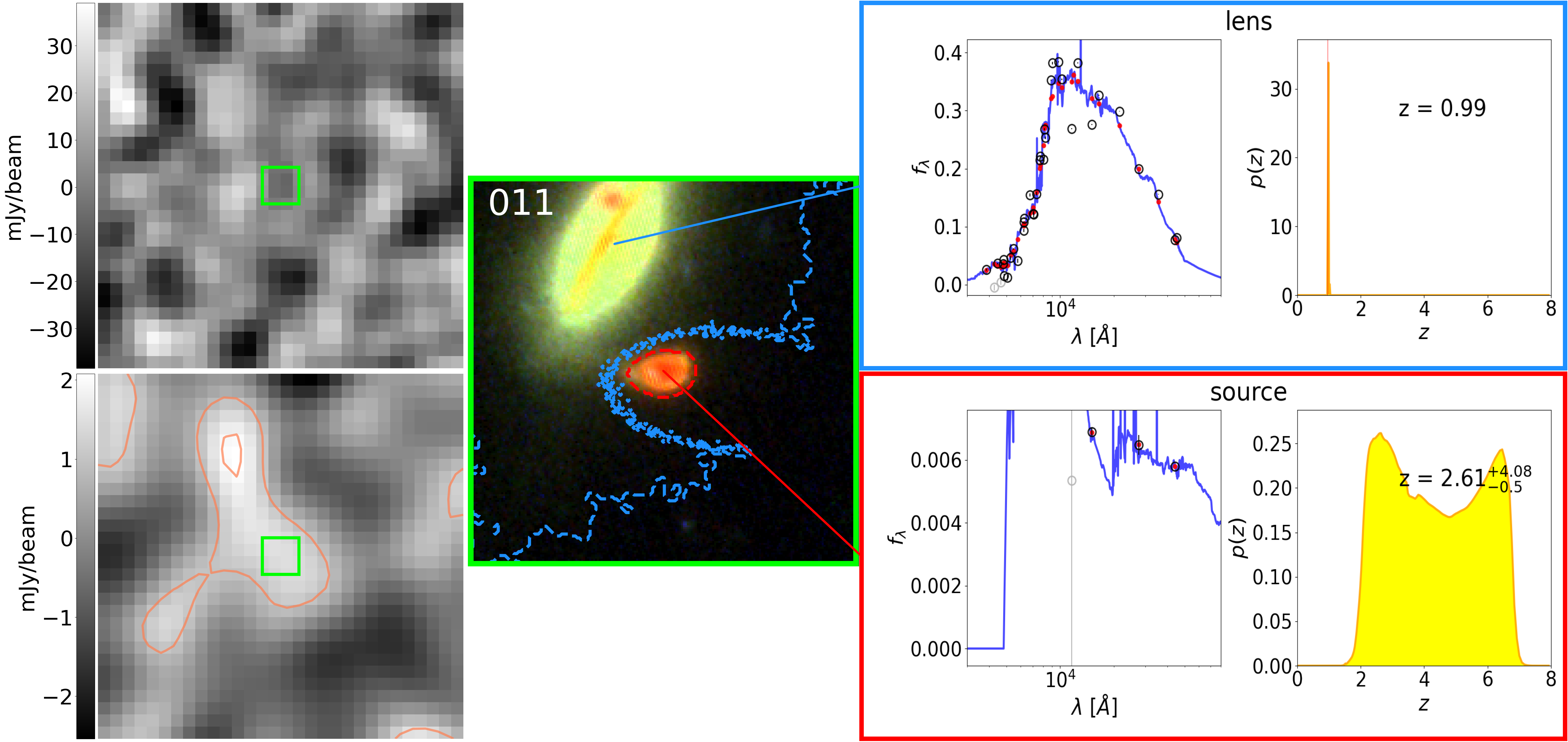} & 
        \includegraphics[width=0.49\textwidth]{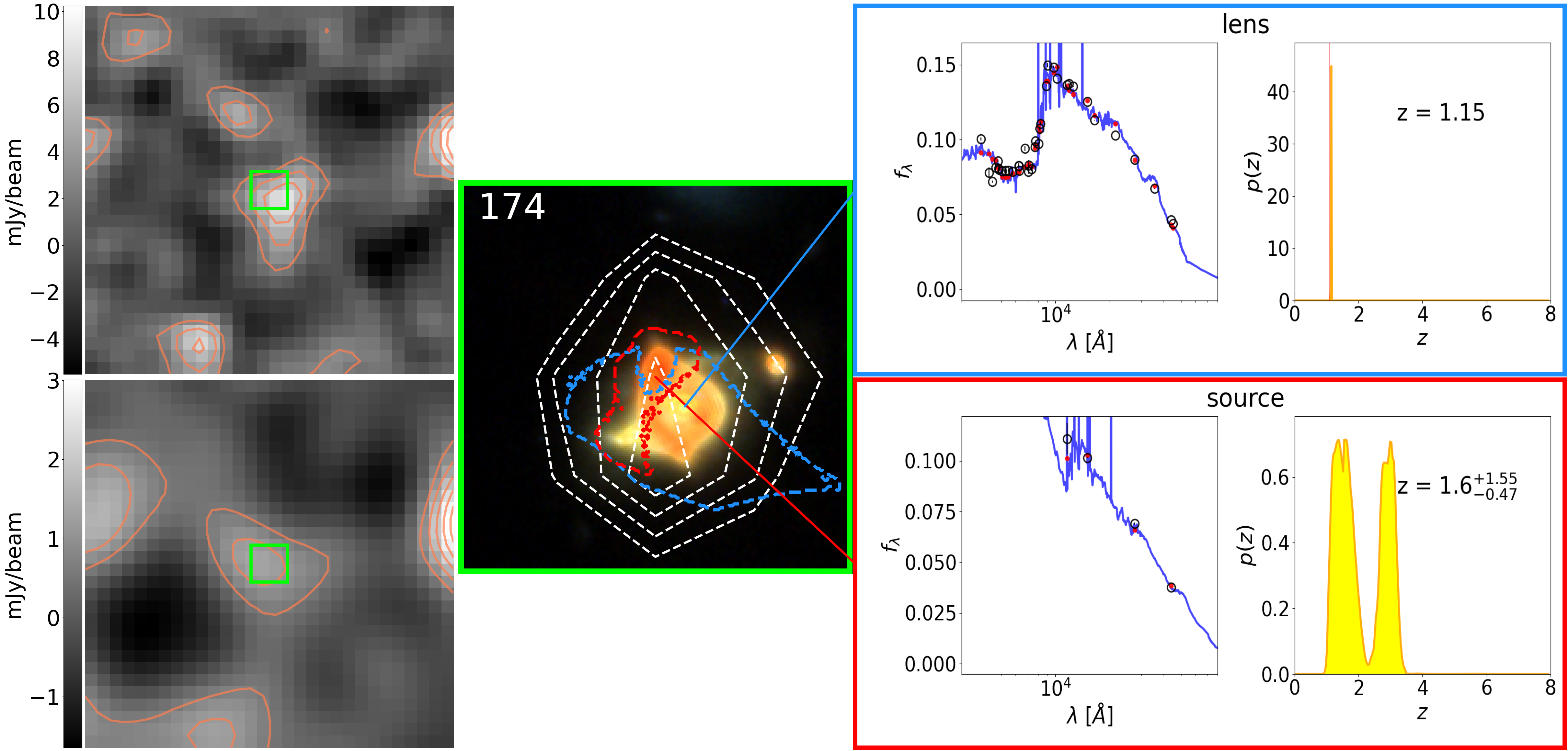} \\     
        \includegraphics[width=0.49\textwidth]{Figure2_cutouts_photz/019.png} & 
        \includegraphics[width=0.49\textwidth]{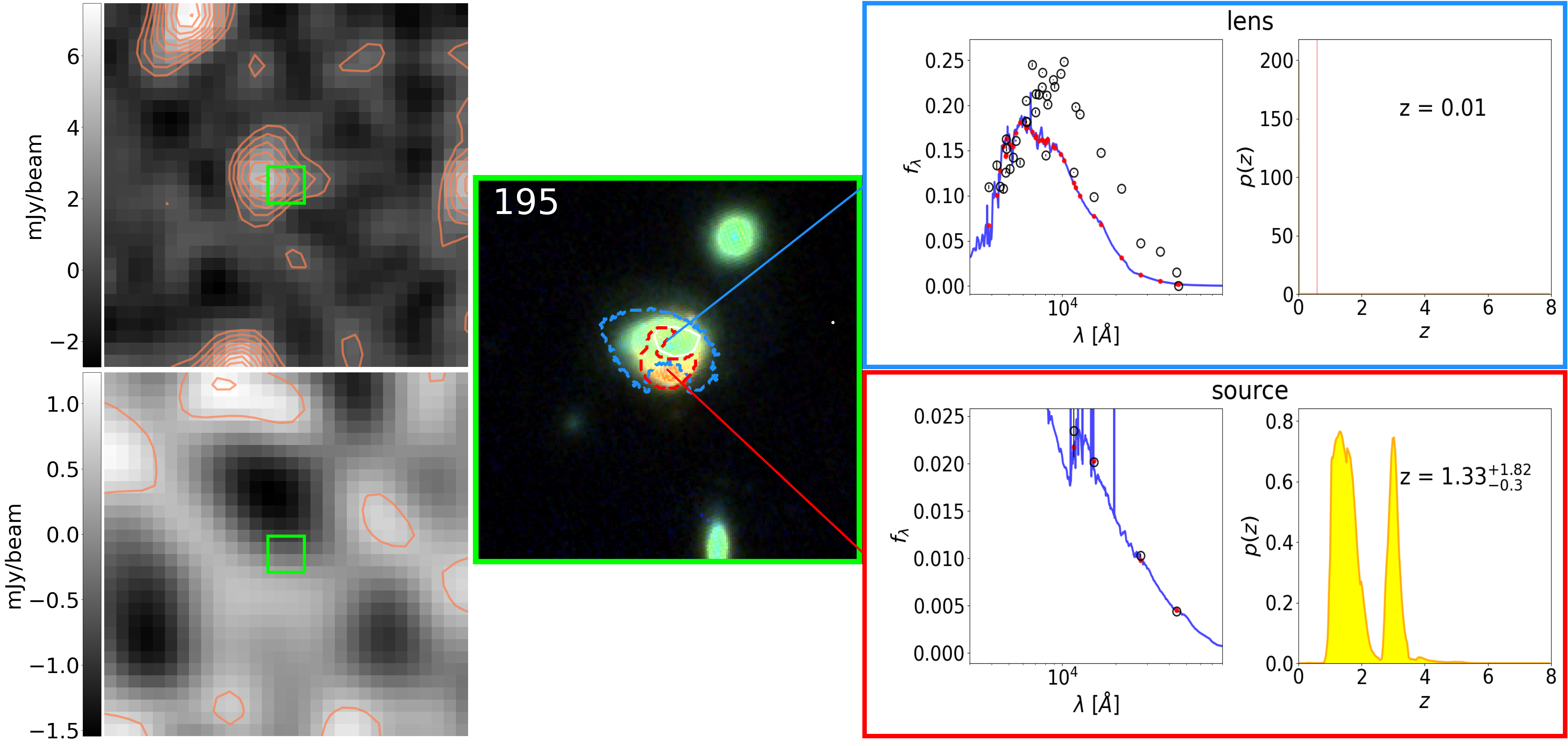} \\ 
        \includegraphics[width=0.49\textwidth]{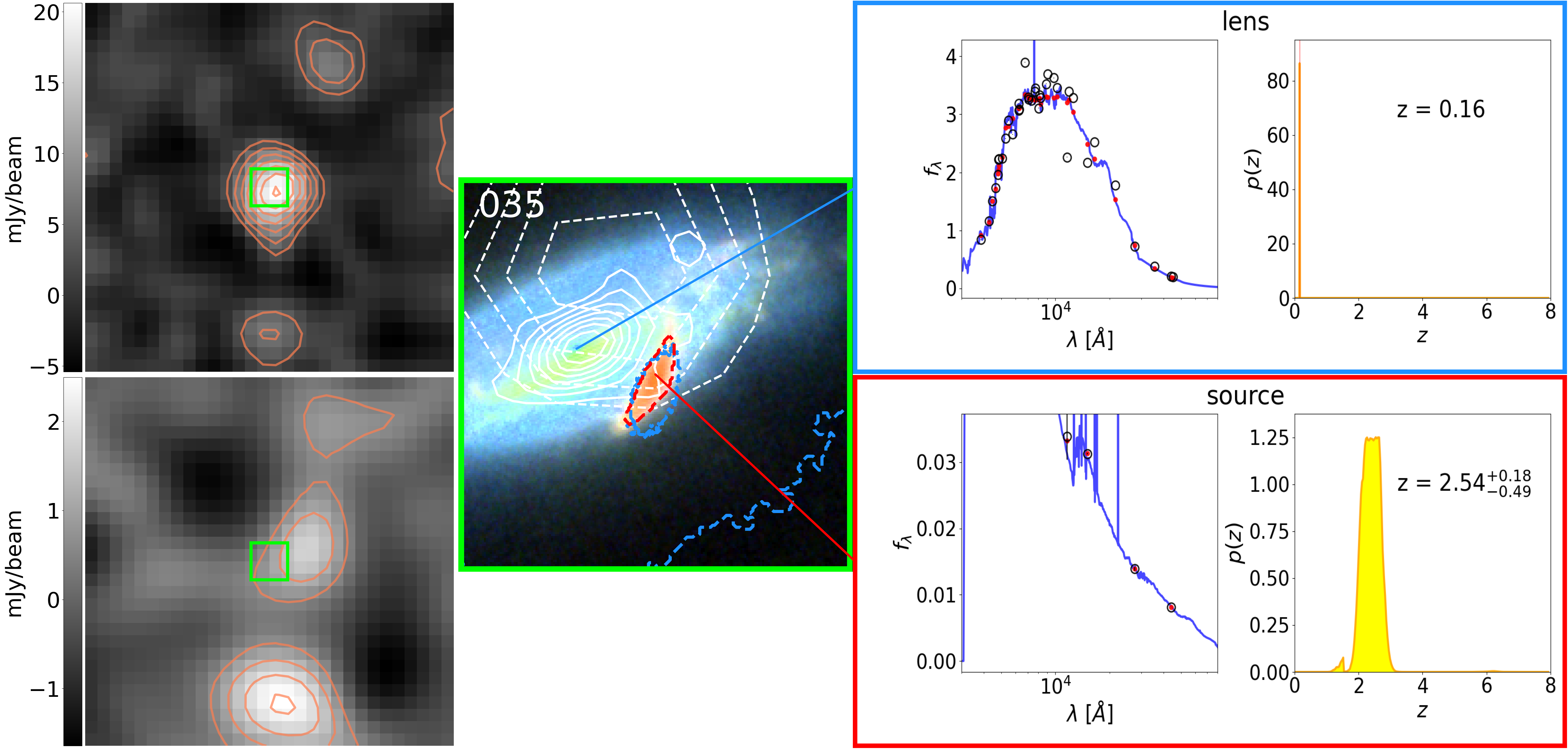} & 
        \includegraphics[width=0.49\textwidth]{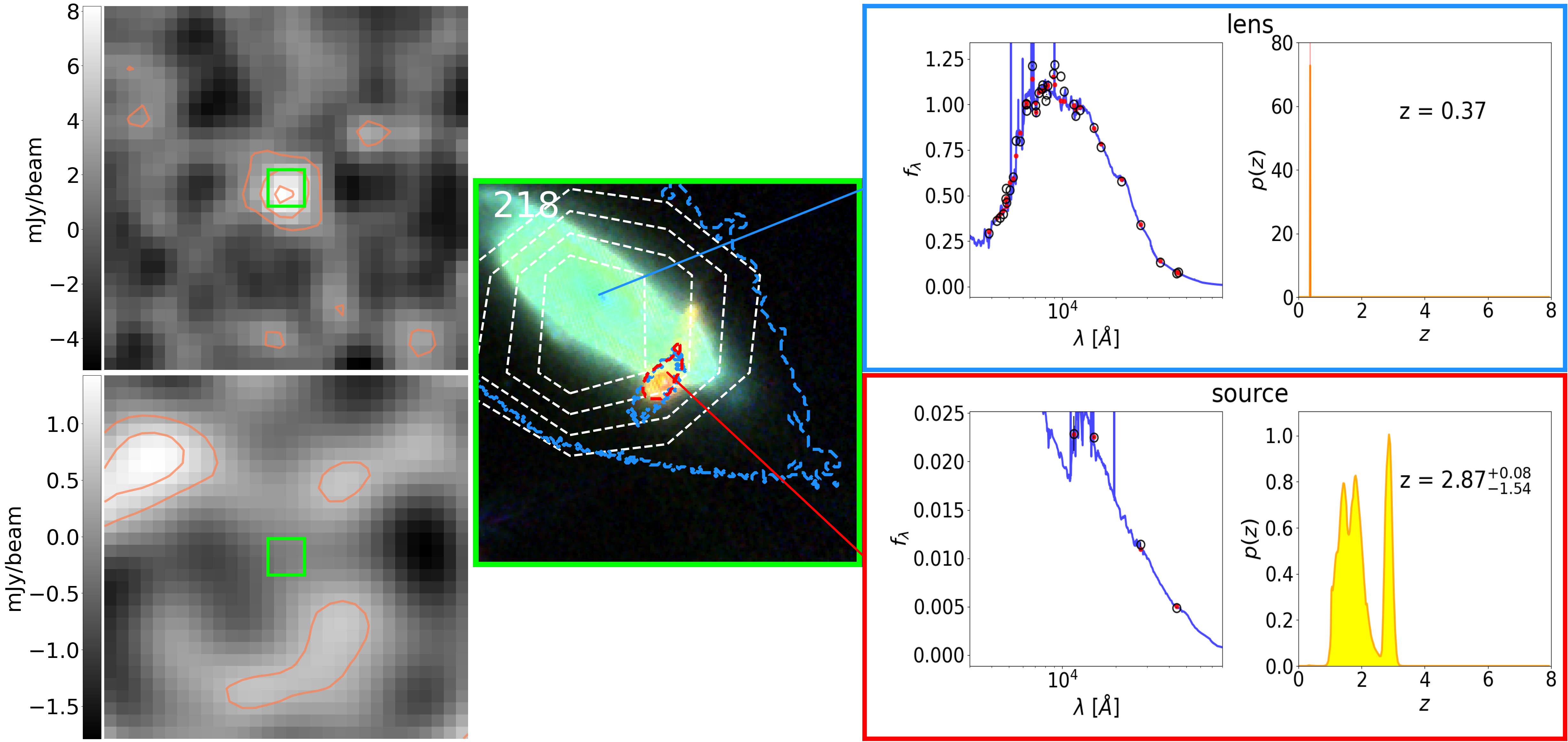} \\ 
        \includegraphics[width=0.49\textwidth]{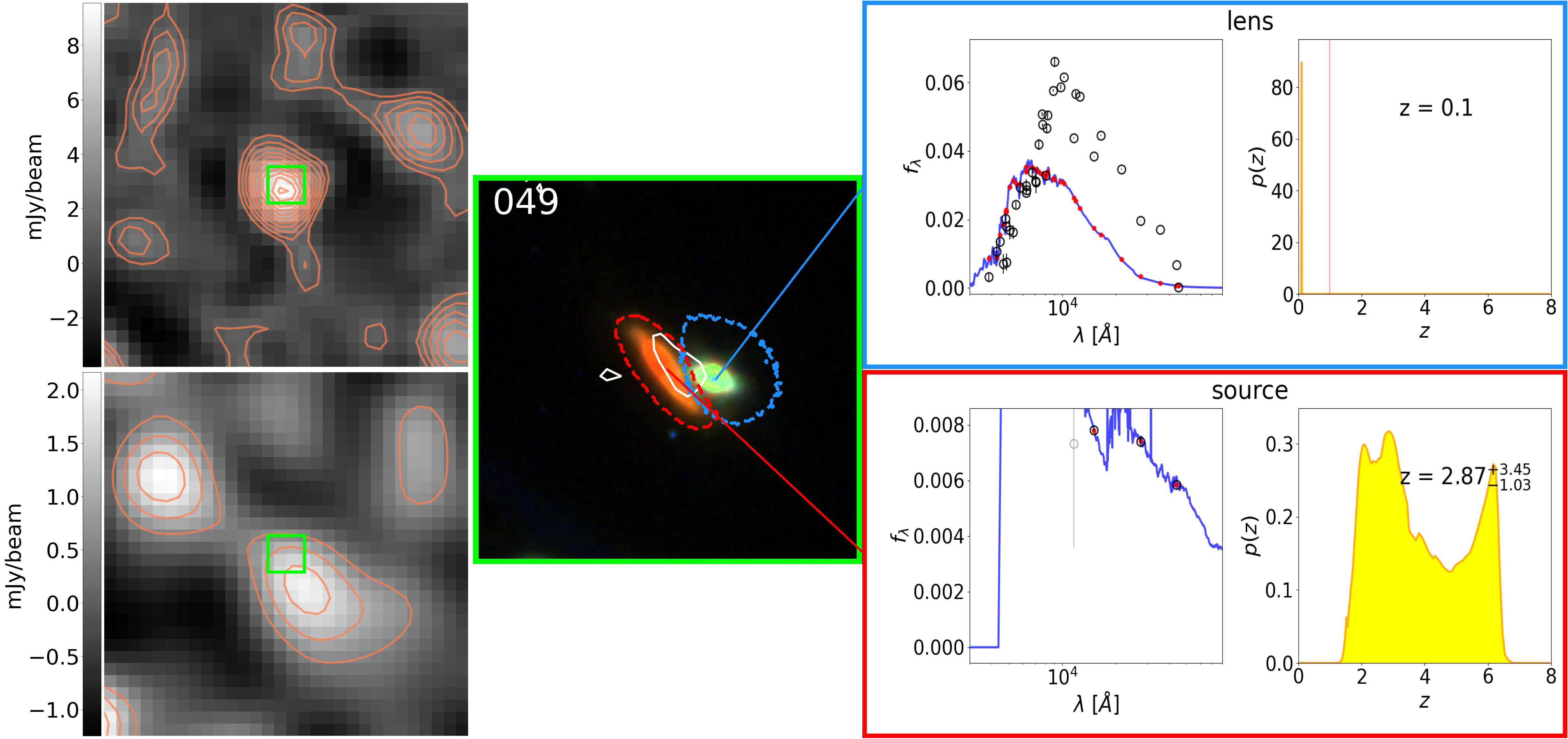} & 
        \includegraphics[width=0.49\textwidth]{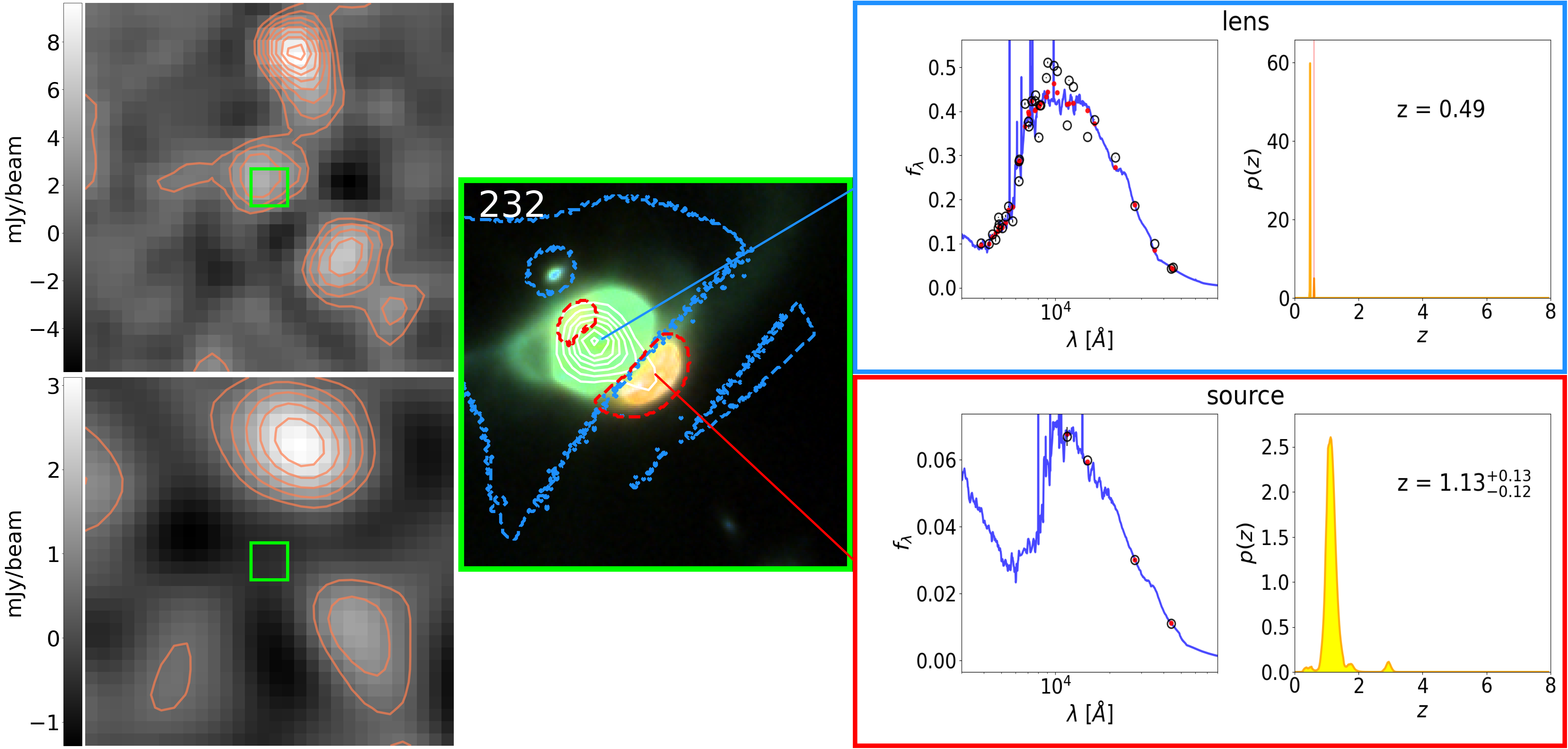} \\ 
        
    \end{tabular}
    \caption{Cutout images and the photo-$z$ analysis for all the 18 candidates, following the style of Figure~\ref{fig:cutout_photz}. For STUDIES sources, the solid contours are either ALMA or VLA, with contour levels starting from 3 with steps of 2$\sigma$. The dashed contours are MIPS emissions, linearly spaced from 80\% to 100\% of the peak flux.}
    \label{fig:A2}
\end{figure}

\begin{figure}[t!]
    \centering
    \begin{tabular}{cc}
    %\centering
        \includegraphics[width=0.49\textwidth]{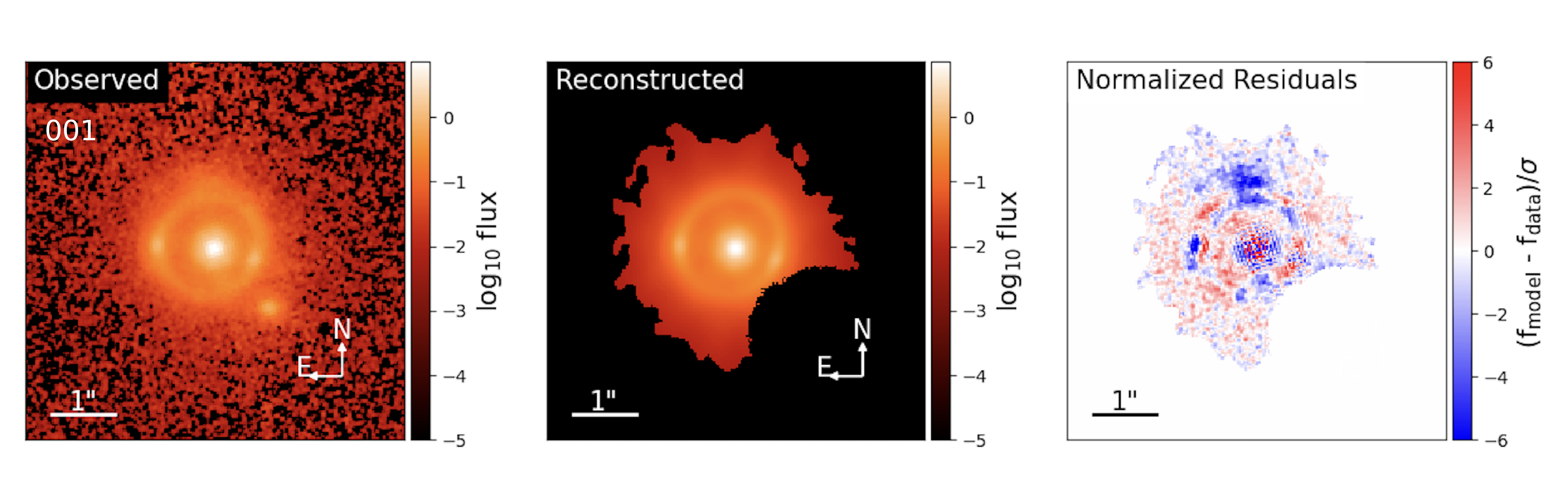} & 
        \includegraphics[width=0.49\textwidth]{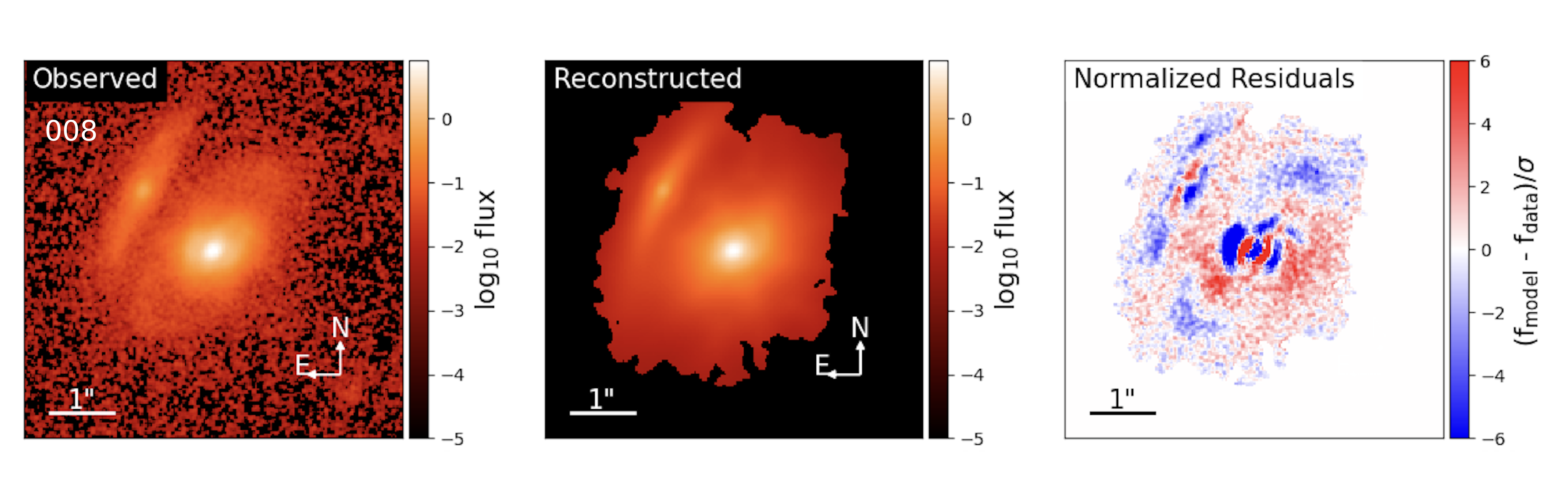} \\     
        \includegraphics[width=0.49\textwidth]{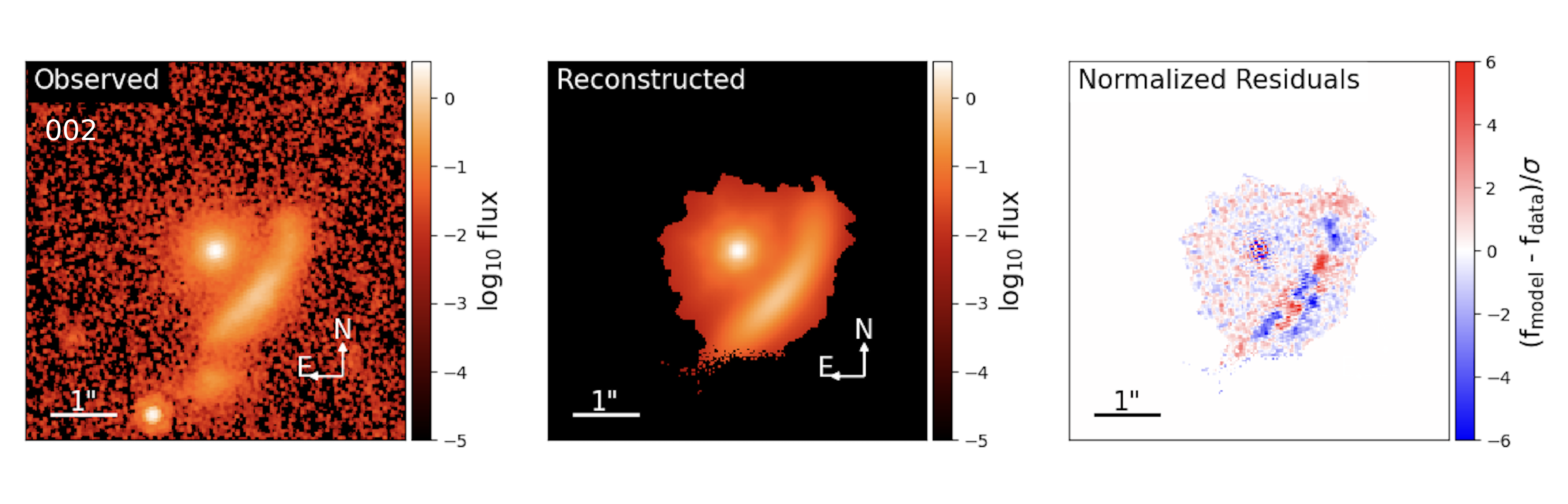} & 
        \includegraphics[width=0.49\textwidth]{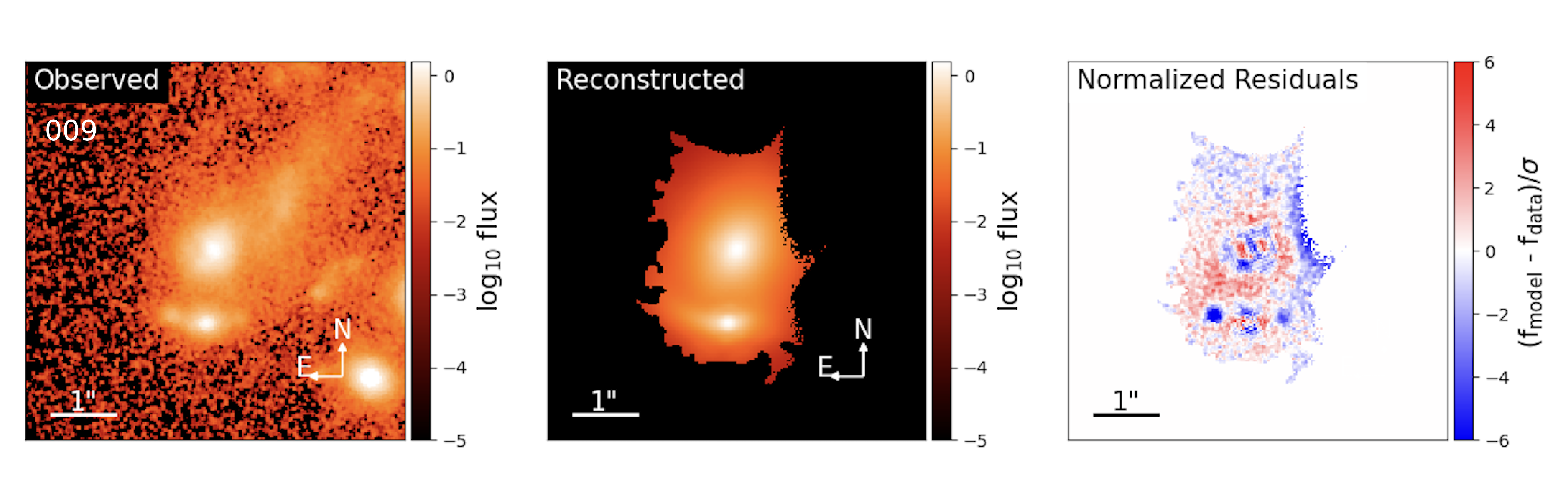} \\    \includegraphics[width=0.49\textwidth]{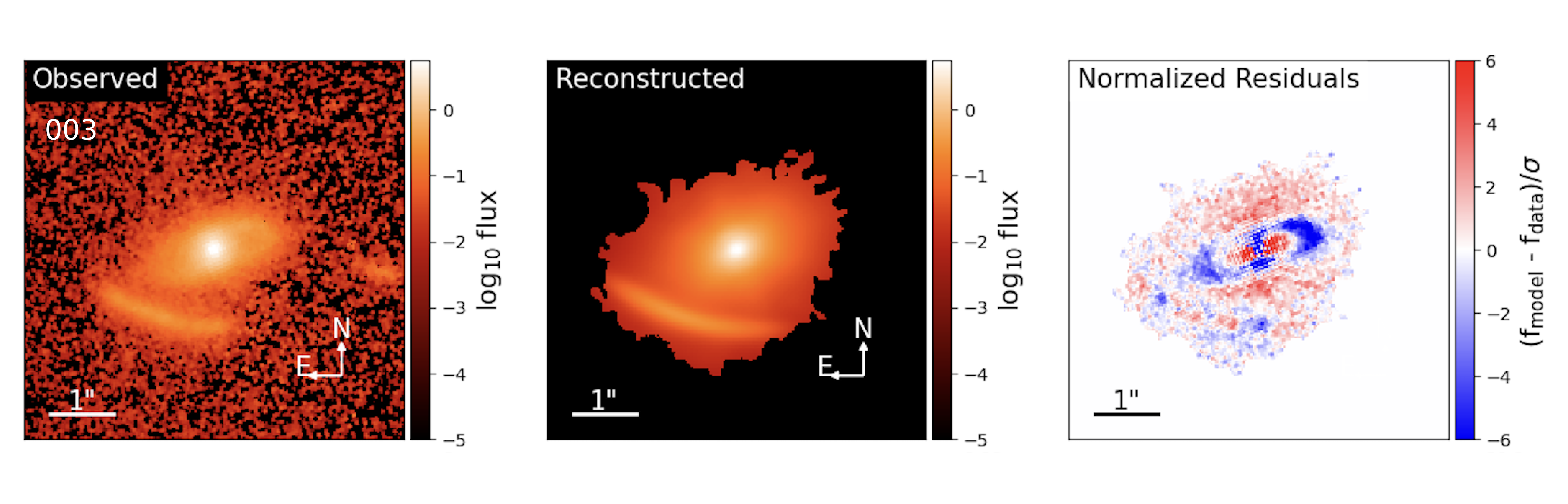} & 
        \includegraphics[width=0.49\textwidth]{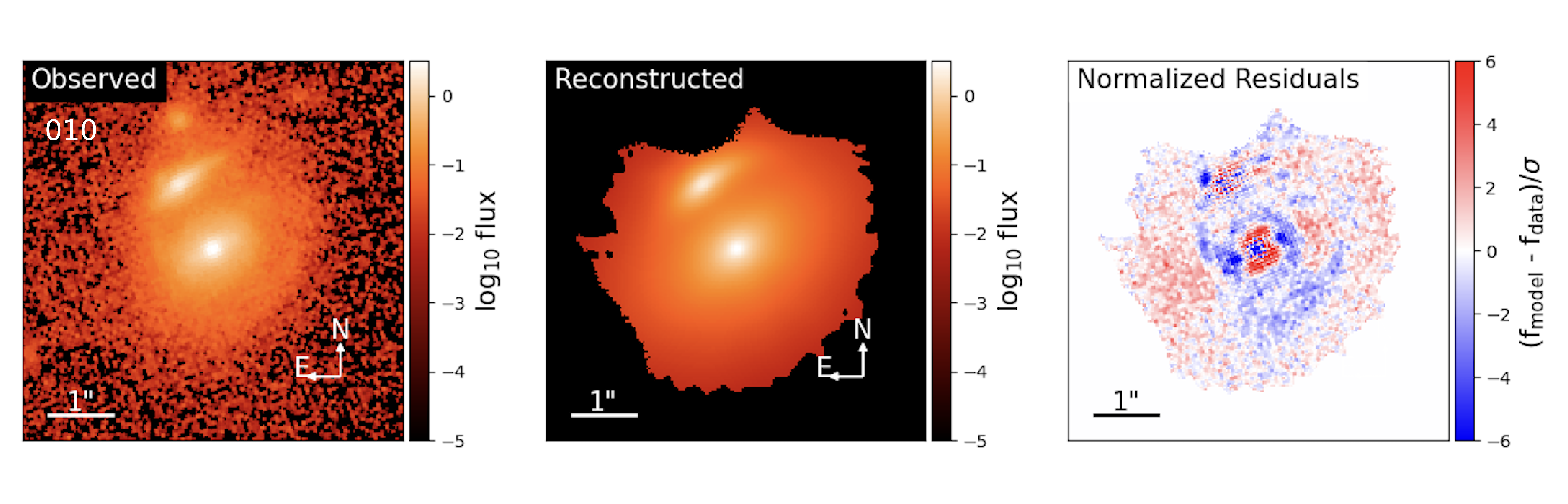} \\    \includegraphics[width=0.49\textwidth]{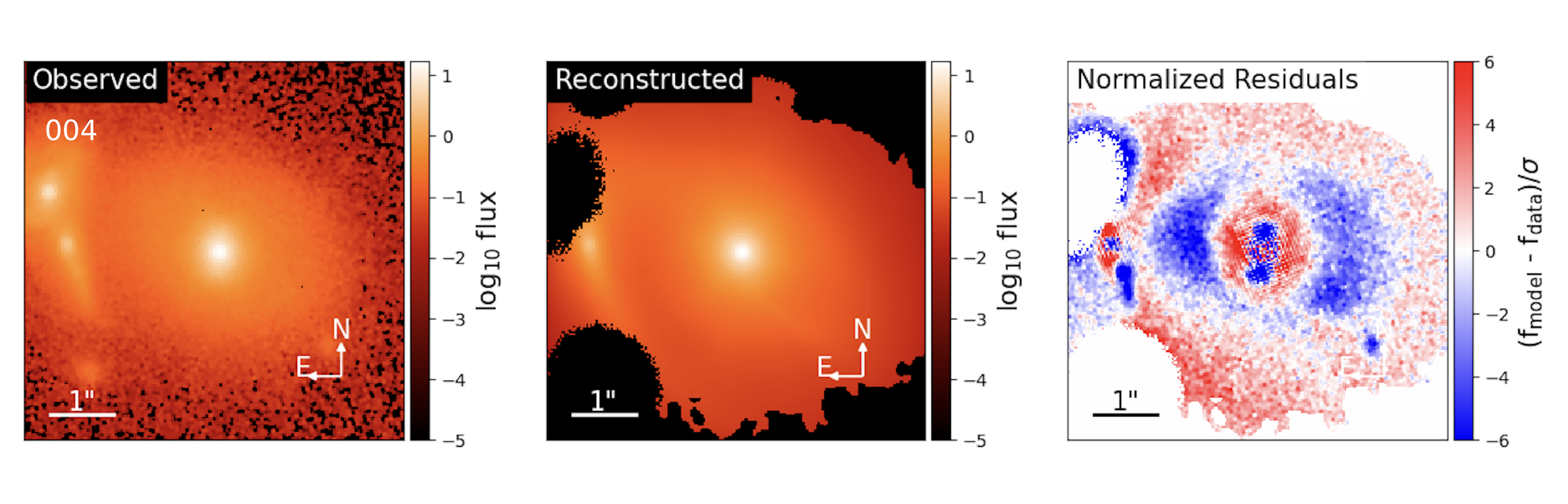} & 
        \includegraphics[width=0.49\textwidth]{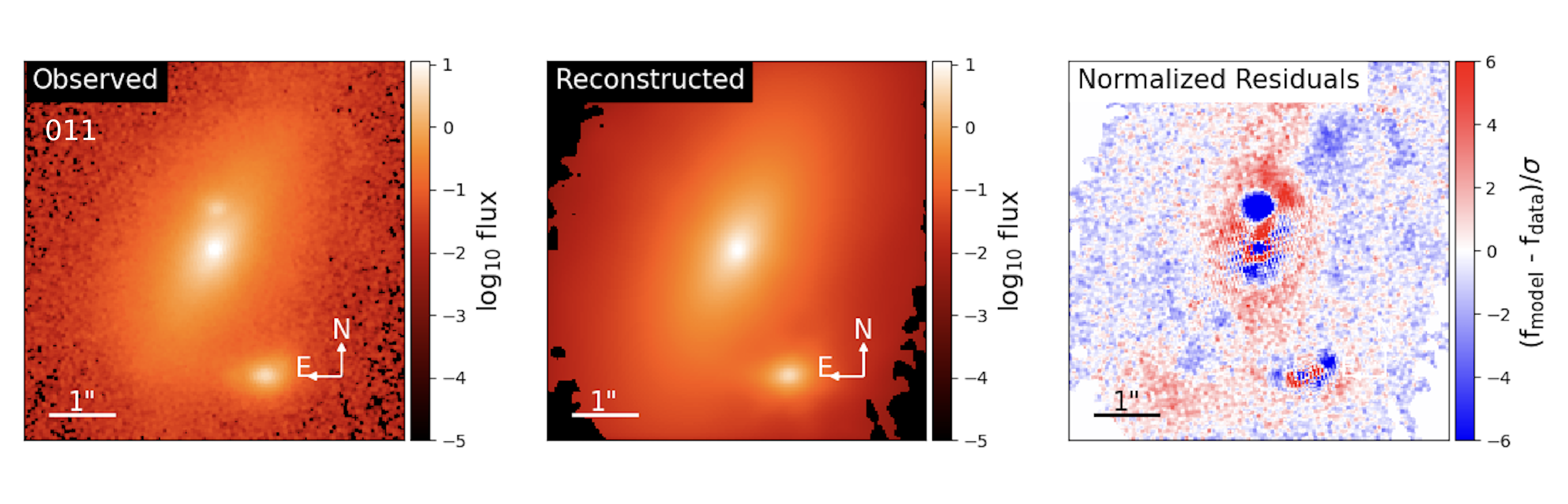} \\    \includegraphics[width=0.49\textwidth]{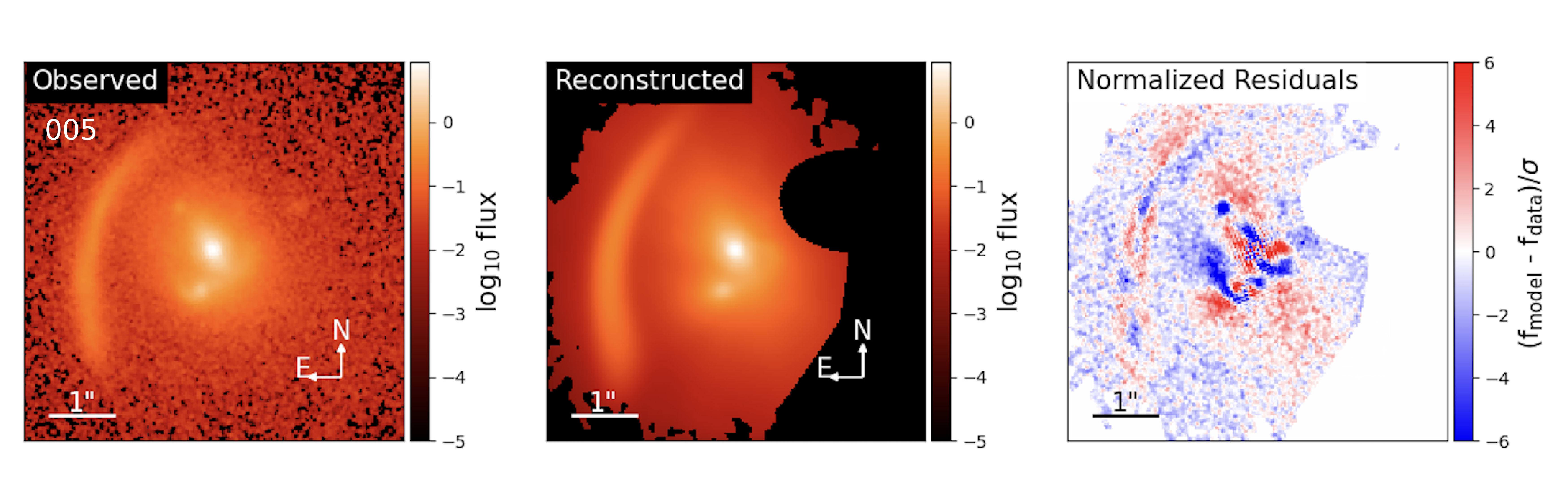} & 
        \includegraphics[width=0.49\textwidth]{Figure3_lens_modeling/019.png} \\    
        \includegraphics[width=0.49\textwidth]{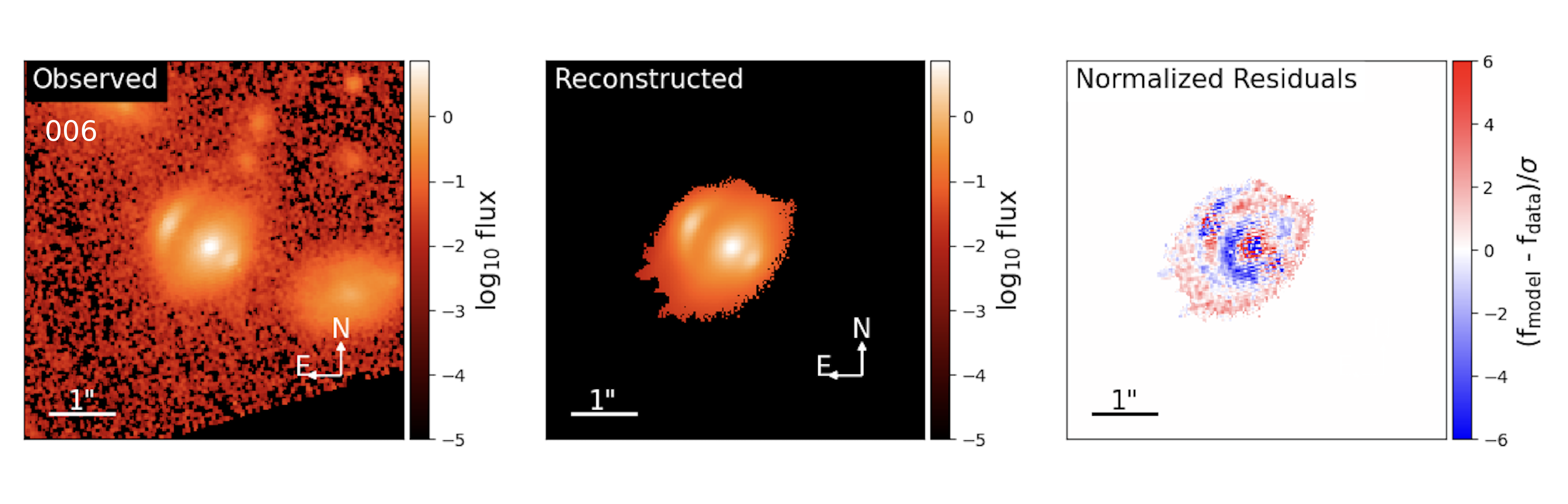} & 
        \includegraphics[width=0.49\textwidth]{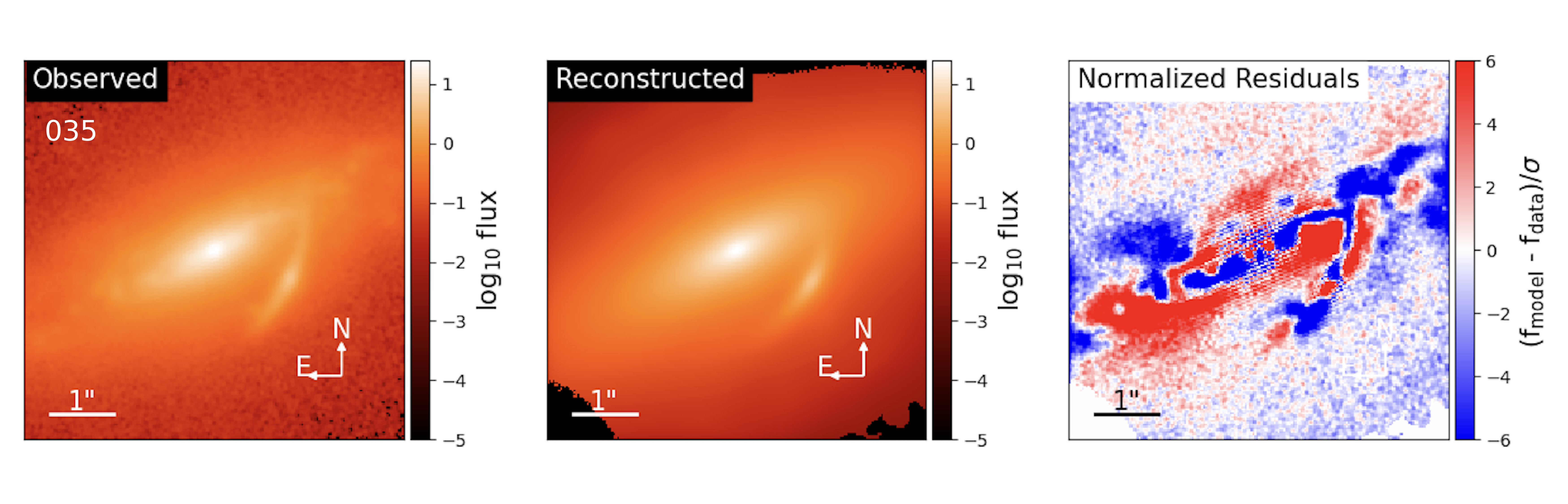} \\  \includegraphics[width=0.49\textwidth]{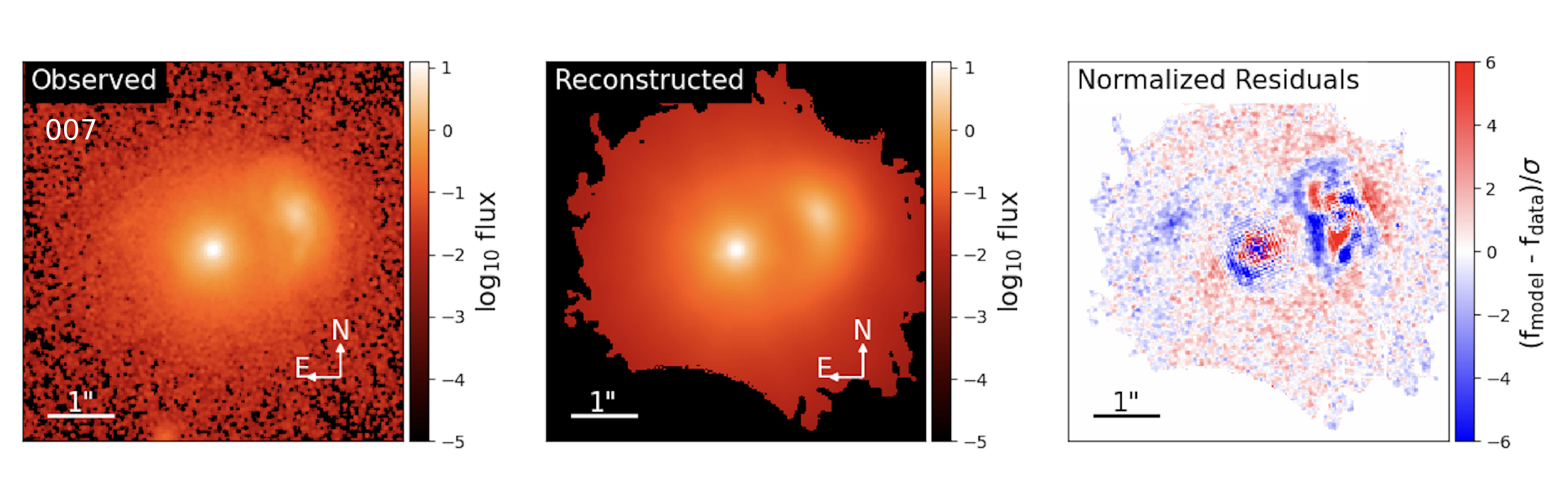} & 
        \includegraphics[width=0.49\textwidth]{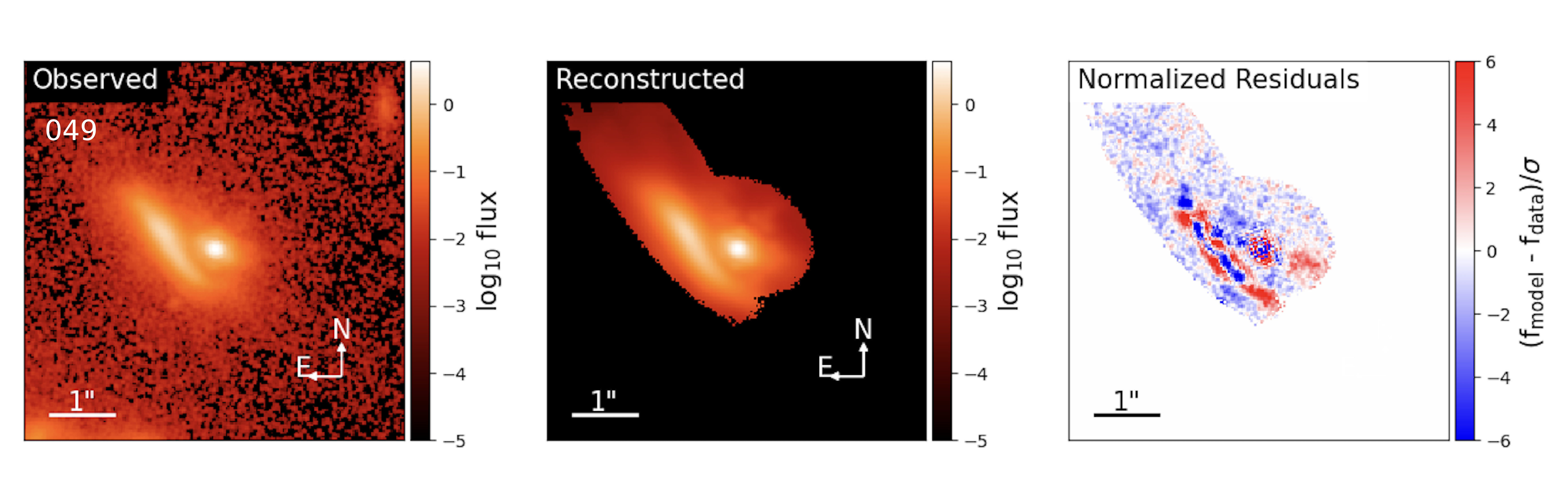} \\  
        
    \end{tabular}
    \caption{Results of the lensing modeling for all the 18 candidates, following the style of Figure~\ref{fig:lens_modeling}.}
    \label{fig:A3}
\end{figure}
\begin{figure}[t!]
    \centering
    \begin{tabular}{cc}
    %\centering
        \includegraphics[width=0.49\textwidth]{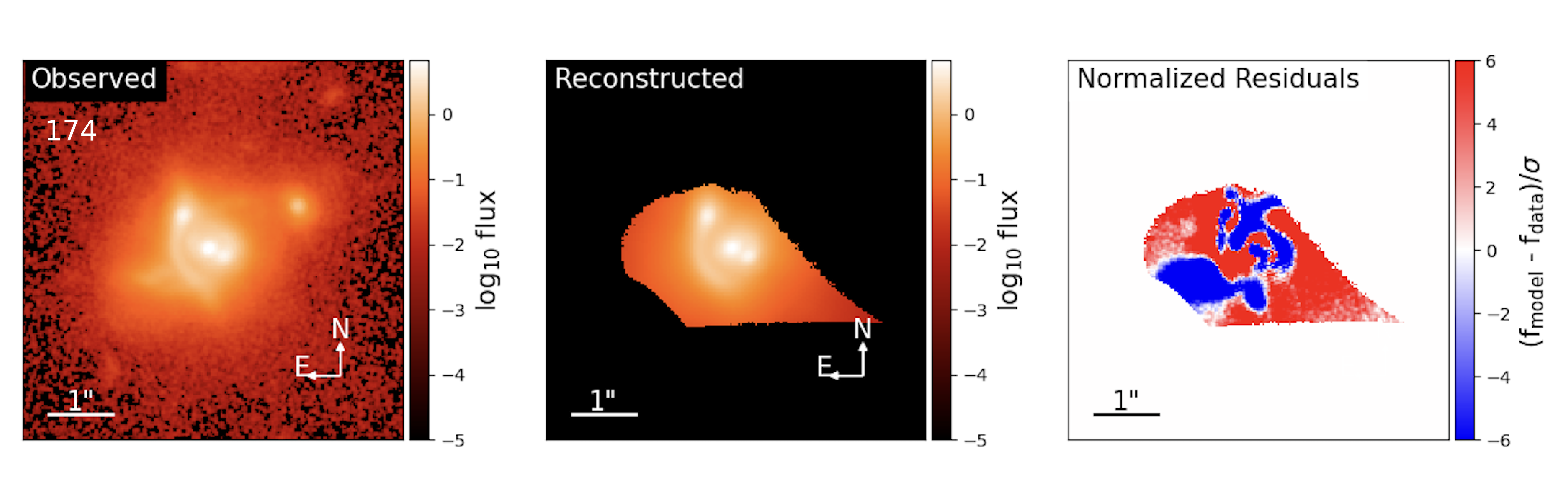} & 
        \includegraphics[width=0.49\textwidth]{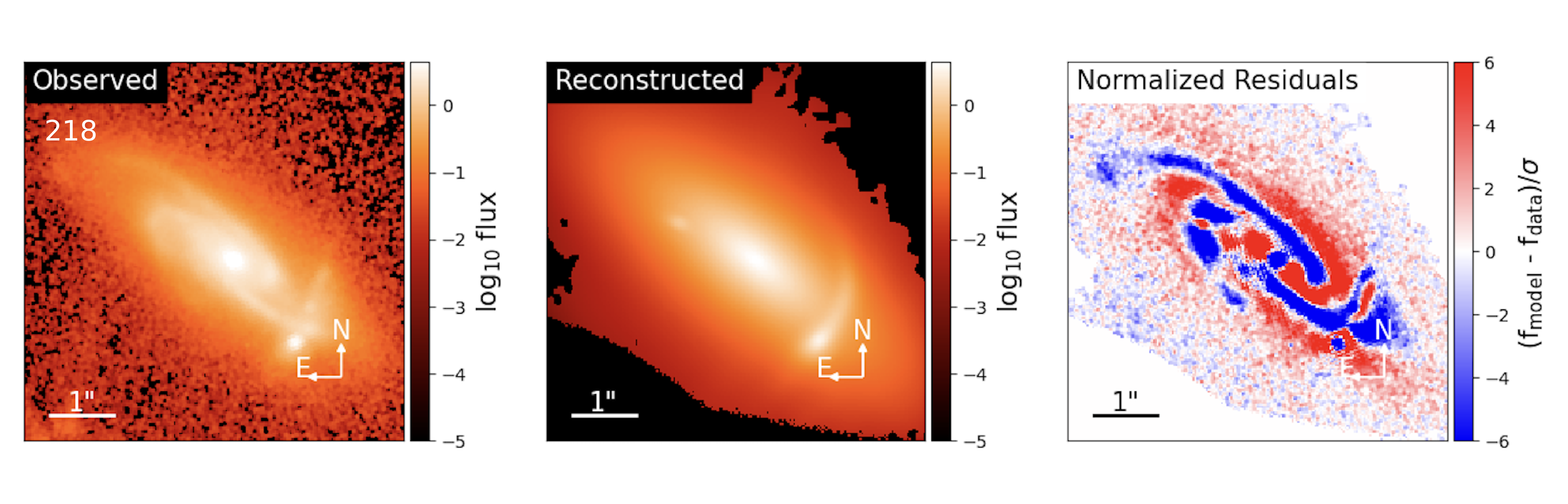} \\     
        \includegraphics[width=0.49\textwidth]{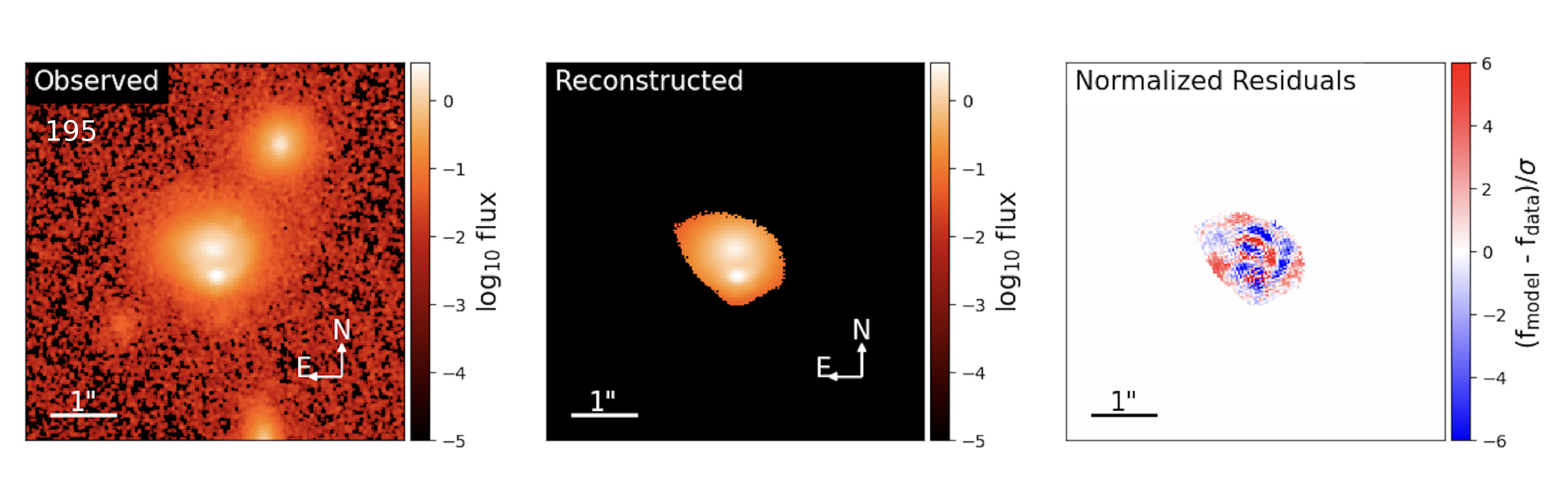} & 
        \includegraphics[width=0.49\textwidth]{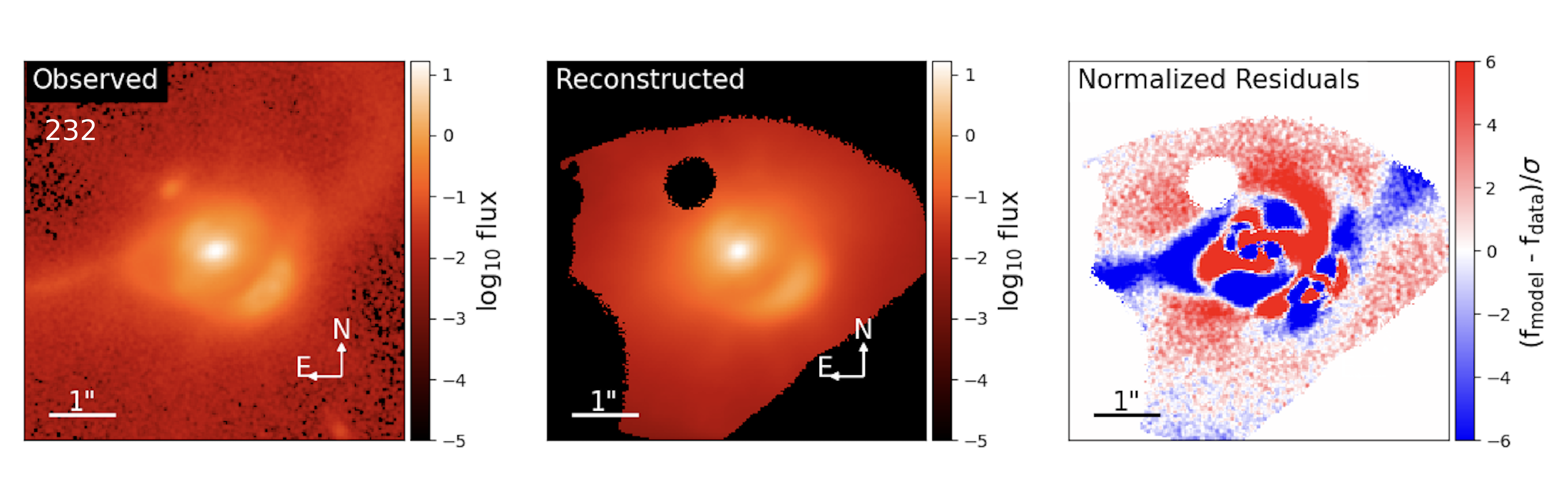} \\    
        
    \end{tabular}
    \caption{Results of the lensing modeling for all the 18 candidates, following the style of Figure~\ref{fig:lens_modeling}.}
    \label{fig:A4}
\end{figure}

\begin{deluxetable}{cc}[h!]
\tablecaption{STUDIES strongly lensed cumulative number count at 450 $\mu$m}

\tablenum{2}

\tablehead{\colhead{S$_{450}$} & \colhead{N($>$S)} \\ 
\colhead{(mJy)} & \colhead{(deg$^{-2}$)} } 

\startdata
18.46 & 8.4$_{-6.9}^{+19.3}$ \\
17.22 & 17.0$_{-14.0}^{+39.0}$ \\
7.50 & 28.4$_{-23.5}^{+65.3}$ \\
6.26 & 42.0$_{-34.7}^{+96.6}$ \\
5.91 & 56.6$_{-46.8}^{+130.2}$ \\
4.23 & 84.2$_{-69.6}^{+193.7}$ \\
3.68 & 121.2$_{-100.2}^{+278.8}$ \\
\enddata

\label{tab:counts}
\end{deluxetable}

\bibliography{sample631}{}
\bibliographystyle{aasjournal}



\end{document}